\documentclass[aos,MSNbibl]{arximspdf}

\pdfoutput = 1

\usepackage{graphicx}

\doi{10.1214/12-AOS1042} 
\volume{40}
\issue{5}
\pubyear{2012}
\firstpage{2572}
\lastpage{2600}

\makeatletter

\newtheorem{theorem}{Theorem}[section]
\newtheorem{proposition}[theorem]{Proposition}

\newtheorem{corollary}[theorem]{Corollary}
\newtheorem{lemma}[theorem]{Lemma}

\newproclaim{algorithm}[theorem]{Algorithm}
\newproclaim{remark}[theorem]{Remark}
\newproclaim{definition}[theorem]{Definition}

\newproclaim{notation}{Notation}

\newcommand{\R}{\mathbb{R}}
\renewcommand{\P}{\mathbb{P}}
\newcommand{\E}{\mathbb{E}}
\newcommand{\cond}{|}

\newcommand{\wh}[1]{{\hat{#1}}}
\newcommand{\eps}{\varepsilon}
\newcommand{\G}{{\widehat{\mathbb{G}}_m}}
\newcommand{\tG}{{\widetilde{\mathbb{G}}_m}}

\newcommand{\BFDR}{\mathrm{BFDR}}
\renewcommand{\l}{\ell}
\newcommand{\thr}{\hat{t}_m}
\renewcommand{\tt}{\tilde{t}_m}

\newcommand{\rate}{\rho_m}

\makeatother

\begin{document}
\begin{frontmatter}

\title{On false discovery rate thresholding for classification under sparsity}
\runtitle{FDR thresholding for classification}

\begin{aug}
\author[A]{\fnms{Pierre} \snm{Neuvial}\ead[label=e1]{pierre.neuvial@genopole.cnrs.fr}}
\and
\author[B]{\fnms{Etienne} \snm{Roquain}\corref{}\thanksref{t1}\ead[label=e2]{etienne.roquain@upmc.fr}}
\runauthor{P. Neuvial and E. Roquain}
\affiliation{Universit\'e d'\'Evry Val d'Essonne and Universit\'{e}
Pierre et Marie Curie}
\address[A]{Laboratoire Statistique et G\'enome\\
Universit\'e d'\'Evry Val d'Essonne\\
UMR CNRS 8071---USC INRA\\
23 boulevard de France\\
91 037 \'Evry\\
France\\
\printead{e1}} 
\address[B]{Laboratoire de Probabilit\'{e}s\\
\quad et Mod\`{e}les Al\'{e}atoires\\
Universit\'{e} Pierre et Marie Curie\\
Paris 6\\
4 Place Jussieu\\
75252 Paris Cedex 05\\
France\\
\printead{e2}}
\end{aug}

\thankstext{t1}{Supported by the French Agence
Nationale de la Recherche (ANR Grants ANR-09-JCJC-0027-01,
ANR-PARCIMONIE, ANR-09-JCJC-0101-01) and by the French ministry of
foreign and European affairs (EGIDE---PROCOPE Project 21887 NJ).}

\received{\smonth{6} \syear{2011}}
\revised{\smonth{8} \syear{2012}}

%
\begin{abstract}
We study the properties of false discovery rate (FDR) thresholding,
viewed as a classification procedure. The ``$0$''-class (null) is
assumed to have a known density while the ``$1$''-class (alternative)
is obtained from the ``$0$''-class either by translation or by
scaling. Furthermore, the ``$1$''-class is assumed to have a small
number of elements w.r.t. the ``$0$''-class (sparsity). We focus on
densities of the Subbotin family, including Gaussian and Laplace
models. Nonasymptotic oracle inequalities are derived for the excess
risk of FDR thresholding. These inequalities lead to explicit rates of
convergence of the excess risk to zero, as the number $m$ of items to
be classified tends to infinity and in a regime where the power of the
Bayes rule is away from $0$ and $1$. Moreover, these theoretical
investigations suggest an explicit choice for the target level
$\alpha_m$ of FDR thresholding, as a function of $m$. Our oracle
inequalities show theoretically that the resulting FDR thresholding
adapts to the unknown sparsity regime contained in the data. This
property is illustrated with numerical experiments.
\end{abstract}

%

\end{frontmatter}

\section{Introduction}\label{secintro}

\subsection{Background}
In many high-dimensional settings, such as microarray or neuro-imaging
data analysis, we aim at detecting signal among several thousands of
items (e.g., genes or voxels).
For such problems, a standard error measure is the false discovery rate
(FDR), which is defined as the expected proportion of errors among the
items declared as significant.

Albeit motivated by pure testing considerations,
the Benjamini--Hochberg FDR controlling procedure~\cite{BH1995} has
recently been shown to enjoy remarkable properties as an estimation
procedure~\cite{ABDJ2006,DJ2006}. More specifically, it turns out to
be adaptive to the amount of signal contained in the data, which has
been referred to as ``adaptation to unknown sparsity.''

In a classification framework, while~\cite{GW2002} contains what is to
our knowledge the first analysis of FDR thresholding with respect to
the mis-classification risk, an important theoretical breakthrough has
recently been made by Bogdan et al.~\cite{BCFG2010}; see also \cite
{BGT2008}. The major contribution of Bogdan et al.~\cite{BCFG2010} is
to create an asymptotic framework in which several multiple testing
procedures can be compared in a sparse Gaussian scale mixture model. In
particular, they proved that FDR thresholding is asymptotically optimal
(as the number $m$ of items goes to infinity) with respect to the
mis-classification risk and thus adapts to unknown sparsity in that
setting (for a suitable choice of the level parameter $\alpha_m$).
Also, they proposed an optimal choice for the rate of $\alpha_m$ as
$m$ grows to infinity.

The present paper can be seen as an extension of~\cite{BCFG2010}.
First, we prove that the property of adaptation to unknown sparsity
also holds nonasymptotically, by using finite sample oracle
inequalities. This leads to a more accurate asymptotic analysis, for
which explicit convergence rates can be provided.
Second, we show that these theoretical properties are not specific to
the Gaussian scale model, but carry over to Subbotin location/scale
models. They can also be extended to (fairly general) log-concave
densities (as shown in the supplemental article~\cite{NR2011supp}),
but we choose to focus on Subbotin densities in the main manuscript for
simplicity.
Finally, we additionally supply an explicit, finite sample, choice of
the level $\alpha_m$ and provide an extensive numerical study that
aims at illustrating graphically the property of adaptation to unknown sparsity.

\subsection{Initial setting}\label{secinsetting}
Let us consider the following classification setting:
let $(X_i,H_i)\in\R\times\{0,1\} $, $1\leq i \leq m$, be $m$ i.i.d.
variables. Assume that the sample $X_1,\ldots,X_m$ is observed without
the labels $H_1,\ldots,H_m$ and that the distribution of $X_1$
conditionally on $H_1=0$ is known a priori.
We consider the following general classification problem: build a
(measurable) classification rule $\hat{h}_m\dvtx\R\rightarrow\{
0,1\}$,
depending on $X_1,\ldots,X_m$, such that the (integrated)
misclassification risk $R_m(\hat{h}_m)$ is as small as possible.
We consider two possible choices for the risk $R_m(\cdot)$:
%
%
\begin{eqnarray}
\label{defrisk-bis-debut} {R}^{T}_m(\hat{h}_m)
&=& \E\Biggl(m^{-1} \sum_{i=1}^m
{\mathbf{1}\bigl\{ \hat{h}_m(X_{i}) \neq H_{i}
\bigr\}} \Biggr);
\\
\label{risk-f-debut} R^{I}_m(\hat{h}_m)&=&\P
\bigl(\hat{h}_m(X_{m+1}) \neq H_{m+1}\bigr),
\end{eqnarray}
where the expectation is taken with respect to $(X_i,H_i)_{1\leq i \leq
m}$ in (\ref{defrisk-bis-debut}) and to $(X_i,H_i)_{1\leq i \leq
m+1}$ in (\ref{risk-f-debut}), for a new labeled data point
$(X_{m+1},H_{m+1}) \sim(X_1,H_1)$ independent\vadjust{\goodbreak} of $(X_i,H_i)_{1\leq i
\leq m}$.
The risks ${R}^{T}_m(\hat{h}_m)$ and $R^{I}_m(\hat{h}_m)$ are usually
referred to as \textit{transductive} and \textit{inductive} risks,
respectively; see Remark~\ref{remrisk} for a short discussion on the
choice of the risk.
Note that these two risks can be different in general because $X_i$
appears ``twice'' in $\hat{h}_m(X_{i})$. However, they coincide for
procedures of the form $\hat{h}_m(\cdot)=h_m(\cdot)$, where
${h}_m\dvtx
\R\rightarrow\{0,1\}$ is a deterministic function. The methodology
investigated here can also be easily extended to a class of \textit
{weighted} mis-classification risks, as originally proposed by Bogdan
et al.~\cite{BCFG2010} (in the case of the transductive risk) and
further discussed in Section~\ref{secweighting}.

The distribution of $(X_1,H_1)$ is assumed to belong to a specific
parametric subset of distributions on $\R\times\{0,1\}$, which is
defined as follows:
\begin{enumerate}[(iii)]
\item[(i)] the distribution of $H_1$ is such that the (unknown)
mixture parameter $\tau_m=\pi_{0,m}/\pi_{1,m}$ satisfies $\tau_m>1$,
where $\pi_{0,m}=\P(H_1=0)$ and $\pi_{1,m}=\P(H_1=1)=1-\pi_{0,m}$;
\item[(ii)] the distribution of $X_1$ conditionally on $H_1=0$ has a
density $d(\cdot)$ w.r.t. the Lebesgue measure on $\R$ that belongs
to the family of so-called $\zeta$-Subbotin densities, parametrized by
$\zeta\geq1$, and defined by
%
%
\begin{eqnarray}
\label{equ-Subdensity}
d(x)&=&(L_\zeta)^{-1} e^{-|x|^\zeta/\zeta}\hspace*{50pt}\nonumber\\[-8pt]\\[-8pt]
&&\eqntext{\displaystyle \mbox{with } L_\zeta= \int_{-\infty}^{+\infty}
e^{-|x|^\zeta/\zeta} \,dx = 2 \Gamma(1/\zeta) \zeta^{1/\zeta-1};}
\end{eqnarray}
\item[(iii)] the distribution of $X_1$ conditionally on $H_1=1$ has a
density $d_{1,m}(\cdot)$ w.r.t. the Lebesgue measure on $\R$ of
either of the following two types:
\begin{enumerate}[-]
\item[-] location: $d_{1,m}(x)=d(x-\mu_m)$, for an (unknown) location
parameter \mbox{$\mu_m>0$};
\item[-] scale: $d_{1,m}(x)\!=\!d(x/\sigma_m)/\sigma_m$, for an
(unknown) scale parameter \mbox{$\sigma_m\!>\!1$}.
\end{enumerate}
\end{enumerate}

\begin{figure}

\includegraphics{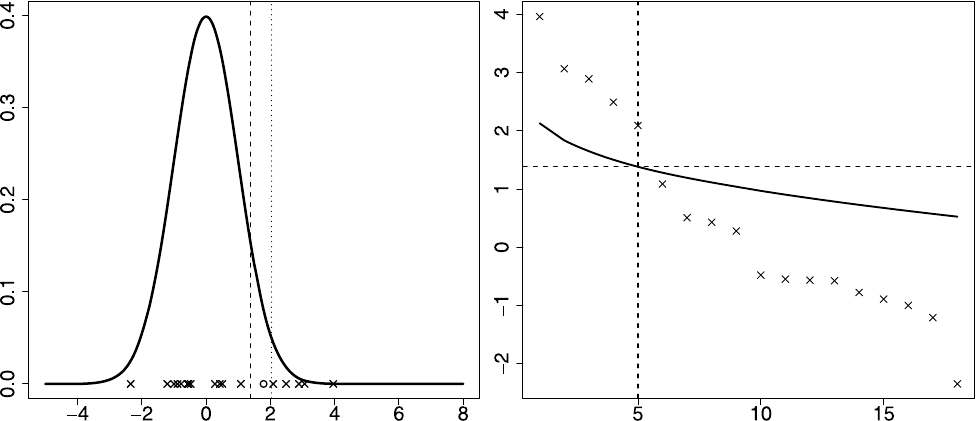}

\caption{Left: illustration of the considered classification problem
for the Gaussian location model for the inductive risk
(\protect\ref{risk-f-debut}); density of $\mathcal{N}(0,1)$ (solid
line); $X_k$,
$k=1,\ldots,m$ (crosses); a new data point $X_{m+1}$ to be classified
(open circle); Bayes' rule (dotted line); FDR rule $\hat{s}^{\mathrm{FDR}}_m$
for $\alpha_m=0.3$ (dashed line). Right: illustration of the FDR
algorithm for $\alpha_m=0.3$; $k\in\{1,\ldots,m\}\mapsto\overline
{\Phi}{}^{-1}(\alpha_m k/m)$ (solid line); $X_{(k)}$'s (crosses);
$\hat
{s}^{\mathrm{FDR}}_m$ (dashed horizontal line); $\hat{k}=5$ (dashed vertical
line). Here, $\overline{\Phi}(x)=\P(X\geq x)$ for $X\sim\mathcal
{N}(0,1)$. $m=18$; $\mu_m=3$; $\tau_m=5$. For this realization, $5$
labels ``$1$'' and $13$ labels ``$0$.''}
\label{figfirstillustration}
\end{figure}

The density $d$ is hence of the form $d(x)=e^{-\phi(|x|)}$ where $\phi
(u)=u^\zeta/\zeta+\log(L_\zeta)$ is convex on $\R^+$ (log-concave
density). This property is of primary interest when applying our
methodology; see the supplemental article~\cite{NR2011supp}.
The particular values $\zeta=1,2$ give rise to the Laplace and
Gaussian case, respectively.
The classification problem under investigation is illustrated by
Figure~\ref{figfirstillustration} (left panel), in the Gaussian
%
%
location case.
Moreover, let us note that we will exclude in our study the Laplace
location model (i.e., the location model using $\zeta=1$). This
particular model is not directly covered by our methodology and needs
specific investigations; see Section 10.3 in the
supplemental article~\cite{NR2011supp}.

Our modeling is motivated by the following application:
consider a microarray experiment for which measurements $Z_1,\ldots
,Z_m$ for $m$ genes are observed, each corresponding to a difference of
expression levels between two experimental conditions (e.g., test
versus reference sample).
Let $H_1,\ldots,H_m$ be binary variables coded as $1$ if the gene is
differentially expressed and $0$ if not. Assume that each $Z_i$ is
$\mathcal{N}(\delta_i,\sigma_\varepsilon^2)$ where $\delta_i$ is
the (unknown) effect for gene $i$ while $\sigma_\varepsilon^2$
quantifies the (known) measurement error. Next, assume the Bayesian
paradigm\vspace*{1pt} that sets the following prior distribution for $\delta_i$:
the distribution of $\delta_i$ is $\mathcal{N}(0,\sigma_0^2)$
conditionally on $H_i=0$ and $\mathcal{N}(\delta,\sigma_0^2 + \tau^2)$
conditionally on $H_i=1$. Generally, $\sigma_0^2$ ($\geq$0), the
dispersion of the nondifferentially expressed genes, is assumed to be
known while $\delta$ ($\geq$0) and $\tau^2$ ($\geq$0), the shift
and additional dispersion of the nondifferentially expressed genes,
are unknown. Let $X_i=Z_i/\sigma$ for $\sigma^2=\sigma_\varepsilon
^2+\sigma_0^2$ and consider the distribution unconditionally on the
$\delta_i$'s. This corresponds to our model (in the Gaussian case) as follows:
\begin{itemize}[-]
\item[-] $\delta>0$ and $\tau^2=0$: location model with $\mu_m=\delta
/\sigma>0$;
\item[-] $\delta=0$ and $\tau^2>0$: scale model with $\sigma_m^2=(\sigma
^2+\tau^2)/\sigma^2>1$.
\end{itemize}
The above convolution argument was originally proposed in \cite
{BCFG2010} for a Gaussian scale model: it explains how we can obtain
test statistics that have the same distribution under the alternative
even if the effects of the measurements are not equal.

Going back to our general setting, an important point is that the
parame\-ters---$(\tau_m,\mu_m)$ in the location model, or $(\tau_m,\sigma
_m)$ in the scale model---are assumed to \textit{depend}
on sample size $m$.
The parameter $\tau_m$, called the \textit{sparsity} parameter, is
assumed to tend to infinity as $m$ tends to infinity, which means that
the unlabeled sample only contains a small, vanishing proportion of
label~$1$. This condition is denoted (\ref{Sp}). As a counterpart, the
other parameter---$\mu_m$ in the location model, or $\sigma_m$ in
the scale model---is assumed to tend to infinity\vadjust{\goodbreak} fast enough to
balance sparsity.
This makes the problem ``just solvable'' under the sparsity constraint.
More precisely, our setting corresponds to the case where the power of
the Bayes procedure is bounded away from $0$ and $1$, and is denoted
(\ref{BP}).
This is motivated by sparse high-dimensional problems for which the
signal is strong but only carried by a small part of the data.
For instance, in the above-mentioned application to microarray data,
the two experimental conditions compared can be so close that only a
very small proportion of genes are truly differentially expressed
(e.g., two groups of patients having the same type of cancer but a
different response to a cancer treatment~\cite{sawyers08the-cancer}).
%
%
\begin{remark}\label{remrisk}
Our setting is close to the semi-supervised novelty detection (SSND)
framework proposed in~\cite{BLS2010}, for which the knowledge of the
distribution $X_1$ conditionally on $H_1=0$ is replaced by the
observation of a finite i.i.d. sample with this distribution.
In the latter work, the authors use the unlabeled data $X_1,\ldots,X_m$
to design a procedure $\wh{h}_m$ that aims at classifying a new
unlabeled data $X_{m+1}$. This approach is in accordance with the
inductive risk defined by (\ref{risk-f-debut}). However, in other
situations closer to standard multiple testing situations, one wants to
classify $X_1,\ldots,X_m$ meanwhile designing $\wh{h}_m$. This gives
rise to the transductive risk defined by (\ref{defrisk-bis-debut}).
\end{remark}

\subsection{Thresholding procedures}\label{secthresholdingproc}

Classically, the solution that minimizes the misclassification risks
(\ref{defrisk-bis-debut}) and (\ref{risk-f-debut}) is the so-called
Bayes rule $h^B_m$ that chooses the label $1$ whenever
$d_{1,m}(x)/d(x)$ is larger than a specific threshold.
We easily check that the likelihood ratio $d_{1,m}(x)/d(x)$ is
nondecreasing in $x$ and $|x|$ for the location and the scale model,
respectively.
As a consequence, we can only focus on classification rules $\hat
{h}_m(x)$ of the form ${\mathbf{1}\{x\geq\hat{s}_m\}}$, $\hat
{s}_m\in\R$
for the location model and ${\mathbf{1}\{|x|\geq\hat{s}_m\}}$, $\hat
{s}_m\in
\R^+$ for the scale model. Therefore, to minimize the
mis-classification risks, thresholding procedures are classification
rules of primary interest, and the main challenge consists of choosing
the threshold $\hat{s}_m$ in function of $X_1,\ldots,X_m$.

The FDR controlling method proposed by Benjamini and Hochberg \cite
{BH1995} (also called ``Benjamini--Hochberg'' thresholding) provides
such a thresholding $\hat{s}_m$ in a very simple way once we can
compute the quantile function $\overline{D}{}^{-1}(\cdot)$, where
$\overline{D}(u)=(L_\zeta)^{-1} \int_u^{+\infty} e^{-|x|^\zeta
/\zeta} \,dx$ is
the (known) upper-tail cumulative distribution function of $X_1$
conditionally on $H_1=0$.
We recall below the algorithm for computing the FDR threshold in the
location model (using test statistics rather than $p$-values).
%
%
\begin{algorithm}\label{algoFDR}
(1) choose a nominal level $\alpha_m\in(0,1)$;

(2) consider the order statistics of the $X_{k}$'s:
$
X_{(1)}\geq X_{(2)} \geq\cdots\geq X_{(m)};
$

(3) take the integer $\hat{k}=\max\{1\leq k \leq m\dvtx
X_{(k)}\geq\overline{D}{}^{-1}(\alpha_m k/m)\}$ when this set is nonempty
and $\hat{k}=1$ otherwise;\vspace*{1pt}

(4) use $\hat{h}^{\mathrm{FDR}}_m(x)={\mathbf{1}\{x\geq\hat
{s}^{\mathrm{FDR}}_m\}}$ for
$\hat{s}^{\mathrm{FDR}}_m=\overline{D}{}^{-1}(\alpha_m
\hat{k}/m)$.\vadjust{\goodbreak}
\end{algorithm}

For the scale model, FDR thresholding has a similar form, $\hat
{h}^{\mathrm{FDR}}_m(x)={\mathbf{1}\{|x|\geq\hat{s}^{\mathrm{FDR}}_m\}}$
for $\hat {s}^{\mathrm{FDR}}_m=\overline{D}{}^{-1}(\alpha_m
\hat{k}/(2m))$,\vspace*{1pt} where $\hat {k}=\max\{1\leq k \leq
m\dvtx\break|X|_{(k)}\geq\overline{D}{}^{-1}(\alpha_m k/ (2m))\}$
($\hat{k}=1$ if the set is empty) and $|X|_{(1)}\geq |X|_{(2)}
\geq\cdots\geq|X|_{(m)}$. Algorithm~\ref{algoFDR} is illustrated in
Figure \ref {figfirstillustration} (right panel), in a Gaussian
location setting. Since
$\hat{s}^{\mathrm{FDR}}_m=\overline{D}{}^{-1}(\alpha_m \hat{k}/m)$
takes its values in the range $[\overline{D}{}^{-1}(\alpha_m),\overline
{D}{}^{-1}(\alpha_m /m)]$, it can be seen as an intermediate
thresholding rule between the Bonferroni thresholding
[$\overline{D}{}^{-1}(\alpha_m /m)$] and the uncorrected thresholding
[$\overline{D}{}^{-1}(\alpha_m)$]. Finally, an important feature of the
FDR procedure is that it depends on a pre-specified level
$\alpha_m\in(0,1)$. In this work, the level $\alpha_m$ is simply used
as a tuning parameter, chosen to make the corresponding
misclassification risk as small as possible. This contrasts with the
standard philosophy of (multiple) testing for which $\alpha_m$ is meant
to be a bound on the error rate and thus is fixed in the overall
setting.

\subsection{Aim and scope of the paper}

Let $R_m(\cdot)$ be the risk defined either by (\ref
{defrisk-bis-debut}) or (\ref{risk-f-debut}).
In this paper, we aim at studying the performance of FDR thresholding
$\hat{h}_m=\hat{h}^{\mathrm{FDR}}_m$ as a classification rule in terms of the
excess risk $R_m(\hat{h}_m)- R_m(h^B_m)$ both in location and scale
models. We investigate two types of theoretical results:
\begin{longlist}[(ii)]
\item[(i)] Nonasymptotic oracle inequalities: prove for each (or
some) $m\geq2$, an inequality of the form
%
%
\begin{equation}
\label{equ-aim1} R_m(\hat{h}_m)- R_m
\bigl(h^B_m\bigr) \leq b_m,
\end{equation}
where $b_m$ is an upper-bound which we aim to be ``as small as
possible.'' Typically, $b_m$ depends on $\zeta,\alpha_m$ and on the
model parameters.
\item[(ii)] Convergence rates: find a sequence $(\alpha_m)_m$ for
which there exists $D>0$ such that for a large $m$,
%
%
\begin{equation}
\label{equ-aim2} R_m(\hat{h}_m)- R_m
\bigl(h^B_m\bigr) \leq D \times R_m
\bigl(h^B_m\bigr) \times\rate
\end{equation}
for a given rate $\rate=o (1 )$.
\end{longlist}

Inequality (\ref{equ-aim1}) is 
of interest in its own right, but
is also used to derive inequalities of type (\ref{equ-aim2}), which
are of asymptotic nature.
Property (\ref{equ-aim2}) is called ``asymptotic optimality at rate
$\rate$.''
It implies that $R_m(\hat{h}_m)\sim R_m(h^B_m)$; that is, $\hat{h}_m$
is ``asymptotically optimal,'' as defined in~\cite{BCFG2010}. However,
(\ref{equ-aim2}) is substantially more informative because it provides
a rate of convergence.

It should be emphasized at this point that the trivial procedure $\hat
{h}^{0}_m\equiv0$ (which always chooses the label ``$0$'') satisfies
(\ref{equ-aim2}) with $\rate=O(1)$ [under our setting (\ref{BP})].
Therefore, proving (\ref{equ-aim2}) with $\rate=O(1)$ is not
sufficient to get an interesting result, and our goal is to obtain a
rate $\rate$ that tends to zero in (\ref{equ-aim2}).
The reason for which $\hat{h}^{0}_m$ is already ``competitive'' is
that we consider a sparse model in which the label ``$0$'' is generated
with high probability.

\subsection{Overview of the paper}

First, Section~\ref{secgenform} presents a more general setting than
the one of Section~\ref{secinsetting}. Namely, the location and scale
models are particular cases of a general ``$p$-value model'' after a
standardization of the original $X_i$'s into $p$-values $p_i$'s.
While the ``test statistic'' formulation is often considered as more
natural than the $p$-value one for many statisticians, the $p$-value
formulation will be very convenient to provide a general answer to our problem.
The so-obtained $p$-values are uniformly distributed on $(0,1)$ under
the label $0$ while they follow a distribution with decreasing density
$f_m$ under the label $1$.
Hence, procedures of primary interest (including the Bayes rule) are
$p$-value thresholding procedures that choose label $1$ for $p$-values
smaller than some threshold $\hat{t}_m$. Throughout the paper, we
focus on this type of procedures, and any procedure $\hat{h}_m$ is
identified by its corresponding threshold $\hat{t}_m$ in the notation.
Translated into this ``$p$-value world,'' we describe in Section \ref
{secgenform} the Bayes rule, the Bayes risk, condition (\ref{BP}),
BFDR and FDR thresholding.

The fundamental results are stated in Section~\ref{secgeneralresults}
in the general $p$-value model.
Following~\cite{ABDJ2006,DJ2004,DJ2006,BCFG2010}, as BFDR thresholding
is much easier to study than FDR thresholding from a mathematical point
of view, the approach advocated here is as follows: first, we state an
oracle inequality for BFDR; see Theorem~\ref{main-th}. Second, we use
a concentration argument of the FDR threshold around the BFDR threshold
to obtain an oracle inequality of the form (\ref{equ-aim1}); see
Theorem~\ref{main-th2}.
At this point, the bounds involve quantities that are not written in an
explicit form, and that depend on the density $f_m$ of the $p$-values
corresponding to the label $1$.

The particular case where $f_m$ comes either from a location or a scale
model is investigated in Section~\ref{secappl-locationandscale}.
An important property is that in these models,
the upper-tail distribution function $\overline{D}(\cdot)$ and the quantile
function $\overline{D}{}^{-1}(\cdot)$ can be suitably bounded; see
Section 12 in the supplemental article~\cite{NR2011supp}.
By using this property, we derive from Theorems~\ref{main-th} and \ref
{main-th2} several inequalities of the form (\ref{equ-aim1}) and
(\ref{equ-aim2}).
In particular, in the sparsity regime $\tau_m=m^{\beta}$, $0<\beta
\leq1$,
we derive that the FDR threshold $\hat{t}^{\mathrm{FDR}}_m$ at level $\alpha_m$
is asymptotically optimal [under (\ref{BP}) and (\ref{Sp})] in
either of the following two cases:
\begin{itemize}[-]
\item[-] for the location model, $\zeta>1$, if $\alpha_m\rightarrow
0$ and $\log\alpha_m=o ( (\log m)^{1-1/\zeta} )$;
\item[-] for the scale model, $\zeta\geq1$, if $\alpha_m\rightarrow
0$ and $\log\alpha_m=o ( \log m )$.
\end{itemize}
The latter is in accordance with the condition found in \cite
{BCFG2010} in the Gaussian scale model.
Furthermore, choosing $\alpha_m \propto1/(\log m)^{1-1/\zeta}$
(location) or $\alpha_m \propto1/(\log m)$ (scale) provides a
convergence rate $\rate=1/ (\log m)^{1-1/\zeta}$ (location) or $\rate
=1/ (\log m)$ (scale), respectively.

At this point, one can argue that the latter convergence results are
not fully satisfactory: first, these results do not provide an explicit
choice for $\alpha_m$ for a given finite value of $m$. Second, the
rate of convergence $\rate$ being rather slow,
we should check numerically that FDR thresholding
has reasonably good performance
for a moderately large $m$.\vadjust{\goodbreak}

We investigate the choice of $\alpha_m$ by carefully studying Bayes'
thresholding and how it is related to BFDR thresholding; see
Sections~\ref{secpFDRthresholding}
and~\ref{secchoicealpham}.
Next, for this choice of $\alpha_m$, the performance of FDR
thresholding is evaluated numerically in terms of (relative) excess
risk, for several values of $m$; see Section~\ref{secnumexp}. We show
that the excess risk of FDR thresholding is small
for a remarkably wide range of values for $\beta$, and increasingly so
as $m$ grows to infinity. This illustrates the adaptation of FDR
thresholding to the unknown sparsity regime.
Also, for comparison, we show that choosing $\alpha_m$ fixed with $m$
(say, $\alpha_m\equiv0.1$) can lead to higher FDR thresholding excess risk.

\section{General setting}\label{secgenform}

\subsection{$p$-value model}\label{secpvaluesetting}

Let $(p_i,H_i)\in[0,1] \times\{0,1\} $, $1\leq i \leq m$, be $m$
i.i.d. variables. The distribution of $(p_1,H_1)$ is assumed to belong
to a specific subset of distributions on $[0,1] \times\{0,1\}$, which
is defined as follows:

\begin{longlist}[(iii)]
\item[(i)] same as (i) in Section~\ref{secinsetting};
\item[(ii)] the distribution of $p_1$ conditionally on $H_1=0$ is
uniform on $(0,1)$;
\item[(iii)] the distribution of $p_1$ conditionally on $H_1=1$ has a
c.d.f. $F_m$ satisfying
{\renewcommand{\theequation}{{A($F_m,\tau_m$)}}
%
\begin{equation}
\label{A}
\begin{tabular}{p{290pt}}
$F_m$
is continuously increasing on $[0,1]$ and differentiable on $(0,1)$,
$f_m=F_m'$ is continuously
decreasing with $f_m(0^+) >\tau_m>
f_m(1^-)$.
\end{tabular}\hspace*{-62pt}
\end{equation}}
\end{longlist}

\noindent This way, we obtain a family of i.i.d. $p$-values, where each
$p$-value has a marginal distribution following the mixture model
\setcounter{equation}{5}
%
\begin{equation}
\label{modelpvalue} p_i \sim\pi_{0,m} U(0,1)+
\pi_{1,m} F_m.
\end{equation}
Model (\ref{modelpvalue}) is classical in the multiple testing
literature and is usually called the ``two-group mixture model.'' It
has been widely used since its introduction by Efron et al. (2001)
\cite{ETST2001}; see, for instance, \cite
{Storey2003,GW2004,Efron2008,BCFG2010}.

The models presented in Section~\ref{secinsetting} are particular
instances of this $p$-value model.
In the scale model, we apply the standardization $p_i=2\overline{D}(|X_i|)$,
which yields $F_m(t)=2\overline{D}(\overline{D}{}^{-1}(t/2)/\sigma_m)$.
In the location model, we let $p_i=\overline{D}(X_i)$, which yields
$F_m(t)=\overline{D}(\overline{D}{}^{-1}(t)-\mu_m)$.
We can easily check that in both cases (\ref{A}) is satisfied
(additionally assuming $\zeta> 1$ for the location model), with
$f_m(0^+)=+\infty$ and $f_m(1^-)<1$ (scale) and $f_m(1^-)=0$
(location), as proved in Section 9.1 in the supplemental
article~\cite{NR2011supp}.

\subsection{Procedures, risks and the Bayes threshold}

A classification procedure is identified with a
threshold $\hat{t}_m\in[0,1]$, that is, a measurable function of the
$p$-value family $(p_i, i\in\{1,\ldots,m\})$. The corresponding procedure
chooses label $1$ whenever the $p$-value is smaller than $\hat{t}_m$.
In the $p$-value setting, the transductive and inductive\vadjust{\goodbreak}
misclassification risks of a threshold $\thr$ can be written as follows:
%
%
\begin{eqnarray}
\label{defrisk-bis} R^T_m(\hat{t}_m)&=&m^{-1}
\sum_{i=1}^m \P(p_i\leq
\hat{t}_m, H_i=0) + m^{-1} \sum
_{i=1}^m \P(p_i> \hat{t}_m,
H_i=1),
\\
\label{defrisk} R^I_m(\hat{t}_m) &=& \E
\bigl( \pi_{0,m} \hat{t}_m + \pi_{1,m}
\bigl(1-F_m(\hat{t}_m)\bigr) \bigr).
\end{eqnarray}
In the particular case of a deterministic threshold $t_m\in[0,1]$,
these two risks coincide and are equal to
$
R_m(t_m)=\pi_{0,m} t_m + \pi_{1,m} (1-F_m(t_m)).
$
The following lemma identifies a solution minimizing both risks (\ref
{defrisk-bis}) and (\ref{defrisk}).
%
%
\begin{lemma}
Let $R_m(\cdot)$ being either $R^T_m(\cdot)$ or $R^I_m(\cdot)$.
Under assumption (\ref{A}), the threshold
%
%
\begin{equation}
\label{thres-Bayes} t^B_{m}=f_m^{-1}(
\tau_m) \in(0,1)
\end{equation}
minimizes $R_m(\cdot)$, that is, satisfies $R_m(t^B_{m})=\min_{\hat
{t}'_m} \{R_m(\hat{t}'_m)\}$,
where the minimum is taken over all measurable functions from $[0,1]^m$
to $[0,1]$ that take as input the $p$-value family $(p_i, i\in\{
1,\ldots,m\})$.
\end{lemma}
The threshold $t^B_m$ is called the \textit{Bayes threshold}, and
$R_m(t^B_m)$ is called the \textit{Bayes risk}.
The Bayes threshold is unknown because it depends on $\tau_m$ and on
the data distribution $f_m$.
%
%
\begin{notation*}
In this paper, all the statements hold for both risks. Hence,
throughout the paper, $R_m(\cdot)$ denotes either $R^T_m(\cdot)$
defined by (\ref{defrisk-bis}) or $R^I_m(\cdot)$ defined by (\ref
{defrisk}).
\end{notation*}

\subsection{Assumptions on the power of the Bayes rule and sparsity}
Under assumption (\ref{A}),
let us denote the power of the Bayes procedure by
%
%
\begin{equation}
\label{equ-powBayes} C_m=F_m\bigl(t^B_{m}
\bigr)\in(0,1).
\end{equation}
In our setting, we will typically assume that the signal is sparse
while the power $C_m$ of the Bayes procedure remains away from $0$ or $1$:
{\renewcommand{\theequation}{BP}
%
\begin{eqnarray}
\label{BP}
&&\mbox{$\exists(C_-,C_+)$ s.t. $\forall m\geq2$,
$0<C_-\leq
C_m \leq C_+<1$};\\[-30pt]\nonumber
\end{eqnarray}}
{\renewcommand{\theequation}{Sp}
%
\begin{eqnarray}
\label{Sp}
&&\mbox{$(\tau_m)_m$ is such that $
\tau_m\rightarrow{+\infty}$ as
$m\rightarrow{+\infty}$}.\\[-30pt]\nonumber
\end{eqnarray}}

\noindent First note that assumption (\ref{Sp}) is very weak: it is
required as
soon as we assume some sparsity in the data. As a typical instance,
$\tau_m=m^\beta$ satisfies (\ref{Sp}), for any $\beta>0$.
Next, assumption (\ref{BP}) means that the best procedure is able to
detect a ``moderate'' amount of signal.
In~\cite{BCFG2010}, a slightly stronger assumption has been introduced,
{\renewcommand{\theequation}{VD}
%
\begin{eqnarray}
\label{VD}
&&\mbox{$\exists C\in(0,1)$ s.t. $C_m\rightarrow C$
as $m$ tends to infinity},\\[-30pt]\nonumber\vadjust{\goodbreak}
\end{eqnarray}}

\noindent which is referred to as ``the verge of detectability.''
Condition (\ref{BP}) encompasses (\ref{VD}) and is more suitable to
state explicit finite sample oracle inequalities; see, for example,
Remark~\ref{remDcomp} further on.

In the location (resp., scale) model, while the original parameters are
$(\mu_m,\tau_m)$ [resp., $(\sigma_m,\tau_m)$],
the model can be parametrized in function of $(C_m,\tau_m)$ by using
(\ref{thres-Bayes}) and (\ref{equ-powBayes}).
This way, $F_m$ is uniquely determined from $(C_m,\tau_m)$ as follows:
among the family of curves $\{\overline{D}(\overline{D}{}^{-1}(\cdot
)-\mu)\}_{\mu
\in\R}$
%
%
\begin{figure}

\includegraphics{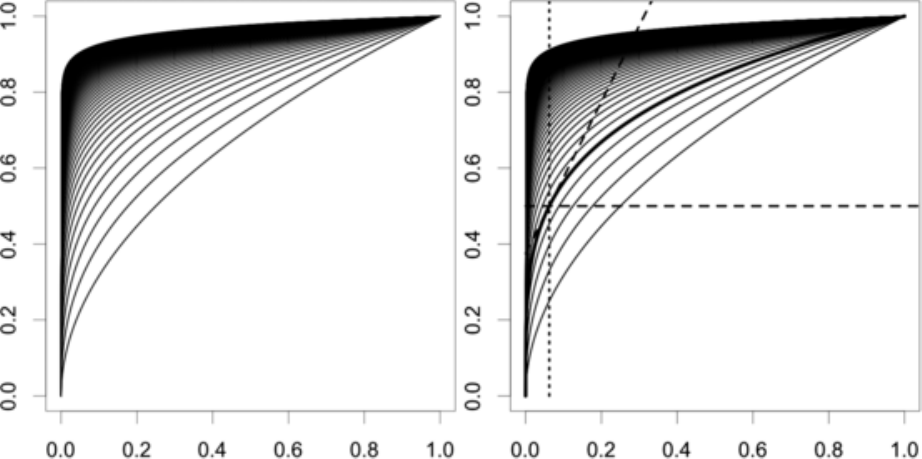}

\caption{Left: plot of the family of curves $\{t\mapsto t^{1/(2+j
/2)}\}_{j=0,\ldots,56}$ (thin solid curves). Right: choice (thick solid
curve) within the family of curves $\{t\mapsto t^{1/\sigma}\}_{\sigma
>1}$ that
fulfills (\protect\ref{thres-Bayes}) and (\protect\ref{equ-powBayes})
for $C_m=1/2$ (given by the dashed horizontal line) and $\tau_m=2$
(slope of the dashed oblique line). This gives $\sigma_m\simeq4$.
The Bayes threshold $t_m^B$ is given by the dotted vertical line.}
\label{figexpo}
\end{figure}
(resp., $\{2\overline{D}(\overline{D}{}^{-1}(\cdot/2)/\sigma)\}_{\sigma>1}$),
$F_m$ is the unique curve such that the pre-image of $C_m$ has a
tangent of slope $\tau_m$, that is, $f_m(F_m^{-1}(C_m))=\tau_m$.
This is illustrated in Figure~\ref{figexpo} for the Laplace scale
model. In this case, $\overline{D}(x)=d(x)=e^{-x}/2$ for $x\geq0$ and thus
$F_m(t)=t^{1/\sigma_m}$, so that the family of curves is simply $\{
t\mapsto t^{1/\sigma}\}_{\sigma>1}$.
%
%
\begin{remark}
Condition (\ref{BP}) constrains the model parameters to be located in
a very specific region.
For instance, in the Gaussian location model with \mbox{$\tau_m=m^\beta$},
condition (\ref{BP}) implies that $\mu_m\sim\sqrt{2\beta\log m}$
(see Table 3 in the supplemental article \cite
{NR2011supp}), which corresponds to choosing $(\mu_m,\beta)$ on the
``estimation boundary,'' as displayed in Figure 1 of~\cite{DJ2004}.
\end{remark}
%
\subsection{BFDR thresholding}\label{secpFDRthresholding}
\mbox{Let us consider the following Bayesian quantity:}
\setcounter{equation}{10}
%
\begin{equation}
\label{BFDR} \BFDR_m(t)=\P(H_i=0\cond p_i
\leq t)=\frac{\pi_{0,m} t}{G_m(t)}= \bigl(1+ \tau_m^{-1}
F_m(t)/t \bigr)^{-1}
\end{equation}
for any $t\in(0,1)$ and where $G_m(t)=\pi_{0,m} t + \pi_{1,m}F_m(t) $.
As introduced by~\cite{ET2002}, the quantity defined by (\ref{BFDR})
is called ``Bayesian FDR.'' It is not to be confounded with ``Bayes
FDR'' defined\vadjust{\goodbreak} by~\cite{SZG2008}.
Also, under a two-class mixture model, $\BFDR_m(t)$ coincides with the
so-called ``positive false discovery rate,''
itself connected to the original false discovery rate of \cite
{BH1995}; see~\cite{Storey2003} and Section 4 of~\cite{BCFG2010}.

Under assumption (\ref{A}), the function $\Psi_m\dvtx t\in
(0,1)\mapsto
F_m(t)/t$ is decreasing from $f_m(0^+)$ to $1$, with $f_m(0^+)\in
(1,+\infty]$. Hence, the following result holds.
%
%
\begin{lemma}\label{lem-pFDR}
Assume (\ref{A}) and $\alpha_m\in((1+f_m(0^+)/\tau_m)^{-1},\pi_{0,m})$.
Then, equation $\BFDR_m(t)= \alpha_m$ has a unique solution
$t=t_m^\star(\alpha_m)\in(0,1)$, given by
%
%
\begin{equation}
\label{thres-pFDR} t^\star_{m}(\alpha_m) =
\Psi_m^{-1} (q_m \tau_m )
\end{equation}
for $q_m=\alpha_m^{-1}-1>0$ and $\Psi_m(t)= F_m(t)/t$.
\end{lemma}
The threshold $t^\star_{m}(\alpha_m)$ is called the \textit{BFDR
threshold} at level $\alpha_m$.
Typically, it is well defined for any $\alpha_m\in(0,1/2)$, because
$\pi_{0,m}>1/2$ and $f_m(0^+)=+\infty$\setcounter{footnote}{1}\footnote{This
condition
implies that the setting is ``noncritical,'' as defined in \cite
{Chi2007}.} in the Subbotin location and scale models (additionally
assuming $\zeta> 1$ for the location model). Obviously, the BFDR
threshold is unknown because it depends on $\tau_m$ and on the
distribution of the data.
However, its interest lies in that it is close to the FDR threshold
which is observable.
When not ambiguous, $t^\star_{m}(\alpha_m)$ will be denoted by $t^\star_{m}$
for short.

Next, a quantity of interest in Lemma~\ref{lem-pFDR} is $q_m=\alpha
_m^{-1}-1 >0$, called the \textit{recovery parameter} (associated to
$\alpha_m$).
As $\alpha_m=(1+q_m)^{-1}$, considering $\alpha_m$ or $q_m$ is equivalent.
Since we would like to have $t^\star_{m}= \Psi_m^{-1}(q_m \tau_m) $
close to $t^B_{m}=f_m^{-1}(\tau_m)$, \textit{the recovery parameter
can be interpreted as a correction factor that cancels the difference
between $\Psi_m(t)=F_m(t)/t$ and $f_m(t)=F_m'(t)$}.
Clearly, the best choice for the recovery parameter is such that
$t^\star_{m}=t^B_{m}$, that is,
%
%
\begin{equation}
\label{qmopt} q_m^{\mathrm{opt}}=\tau_m^{-1}
\Psi_m\bigl(f_m^{-1}(\tau_m)\bigr) =
\frac
{C_m}{\tau_m t_m^B},
\end{equation}
which is an unknown quantity, called the \textit{optimal recovery
parameter}. Note that from the concavity of $F_m$, we have $\Psi
_m(t)\geq f_m(t)$ and thus $q_m^{\mathrm{opt}}\geq1$.
As an illustration, for the Laplace scale model, we have $\sigma_m
f_m(t)=\Psi_m(t)$ and thus the optimal recovery parameter is
$q_m^{\mathrm{opt}}
=\sigma_m$.

The fact that $q_m^{\mathrm{opt}}\geq1$ suggests to always choose
$q_m\geq1$
(i.e., $\alpha_m\leq1/2$) into the BFDR threshold. A related result
is that taking any sequence $(\alpha_m)_m$ such that $\alpha_m \geq
\alpha_-> 1/2$ for all $m\geq2$ never leads to an asymptotically
optimal BFDR procedure; see Section 13 in the
supplemental article~\cite{NR2011supp}.

%
%
\begin{figure}

\includegraphics{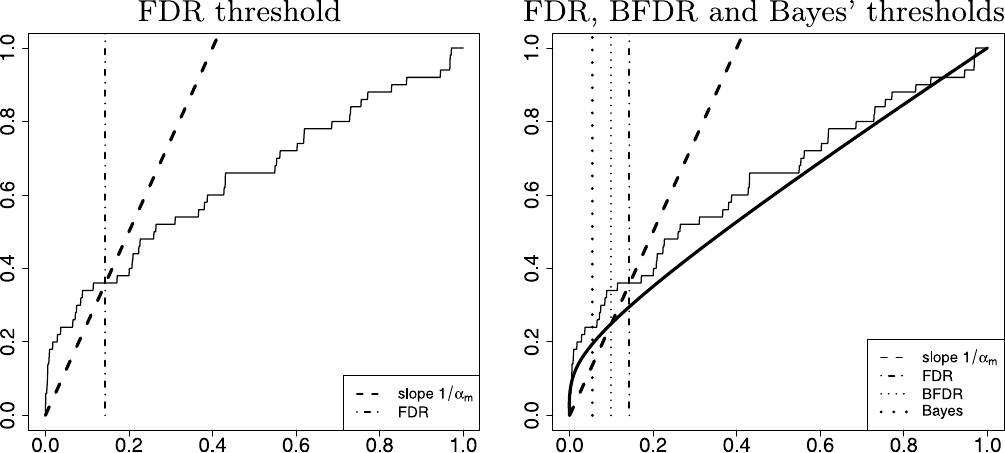}

\caption{Left: illustration of the FDR threshold
(\protect\ref{equ-FDRsupBonf}): e.c.d.f. of the $p$-value (solid
line), line of slope
$1/\alpha_m$ (dotted line), the FDR threshold at level $\alpha_m$
($X$-coordinate of the vertical dashed dotted line). Right:
illustration of the FDR threshold as an empirical surrogate for the
BFDR threshold; compared to the left picture, we added the c.d.f. of
the $p$-values (thick solid line), the BFDR threshold at level $\alpha
_m\pi_{0,m}$ (dotted vertical line) and the Bayes threshold (dashed
vertical line).
In both panels, we consider the Laplace scale model with $C_m=0.5$;
$m=50$; $\beta=0.2$; $\tau_m=m^\beta$; $\sigma_m\simeq4.2$;
$\alpha_m=0.4$.}
\label{figexpoFDR}
\end{figure}
%

\subsection{FDR thresholding}


The FDR threholding procedure was introduced by Benjamini and Hochberg
(1995) by proving that it controls the FDR; see~\cite{BH1995}. From an
historical perspective,\vadjust{\goodbreak} it is interesting to note that this procedure
has a prior occurrence in a series of papers by Eklund (1961--1963); see
\cite{See1968}.
As noted by many authors (see, e.g., \cite
{Sen1999,ET2002,Storey2002,GW2002,ABDJ2006}), this thresholding rule
can be expressed
as a function of the empirical c.d.f. $\G$ of the $p$-values in the
following way: for any $\alpha_m\in(0,1)$,
%
%
\begin{equation}
\label{equFDR} \hat{t}^{\mathrm{BH}}_m(\alpha_m) = \max
\bigl\{t\in[0,1]\dvtx\G(t)\geq t/\alpha_m \bigr\}.
\end{equation}
We simply\vspace*{2pt} denote $\hat{t}^{\mathrm{BH}}_m(\alpha_m)$ by $\hat{t}^{\mathrm{BH}}_m$ when
not ambiguous.
Classically, this implies that $t=\hat{t}^{\mathrm{BH}}_m$ solves the equation
$\G(t)= t/\alpha_m$ [this\vspace*{1pt} can be easily shown by using (\ref
{equFDR}) together with the fact that $\G(\cdot)$ is a nondecreasing
function].
Hence, according to Lemma~\ref{lem-pFDR} and as already mentioned in
the literature (see~\cite{BCFG2010}), $\hat{t}^{\mathrm{BH}}_m$~can be seen as
an empirical counterpart of the BFDR threshold at level $\alpha_m\pi
_{0,m}$, in which the theoretical c.d.f. $G_m(t)=\pi_{0,m}t+\pi_{1,m}
F_m(t)$ of the $p$-values\vspace*{1pt} has been replaced by the empirical c.d.f. $\G
$ of the $p$-values.
Next, once $\alpha_m$ has been chosen, (\ref{equFDR}) only involves
observable quantities, so that the threshold $\hat{t}^{\mathrm{BH}}_m$ only
depends on the data.
This is further illustrated on the left panel of Figure~\ref{figexpoFDR}.
Also, as already observed in Section 5.2 of~\cite{BCFG2010}, since the
BH procedure
is never more conservative than the Bonferroni procedure, the following
modification of $\hat{t}^{\mathrm{BH}}_m$ can be proposed:

%
\begin{definition}
The \textit{FDR threshold} at level $\alpha_m$ is defined by
%
%
\begin{equation}
\label{equ-FDRsupBonf} \hat{t}_m^{\mathrm{FDR}}(\alpha_m)=
\hat{t}^{\mathrm{BH}}_m(\alpha_m) \vee(
\alpha_m/m),
\end{equation}
where $\hat{t}^{\mathrm{BH}}_m(\alpha_m)$ is defined by
(\ref{equFDR}).
\end{definition}
We simply denote $\hat{t}^{\mathrm{FDR}}_m(\alpha_m)$ by
$\hat{t}^{\mathrm{FDR}}_m$ when not ambiguous. The threshold
$\hat{t}^{\mathrm{FDR}}_m$ is the one that we use
throughout\vspace*{2pt} this paper. This modification does not change
the risk $R^T_m(\cdot)$, that is,
$R^T_m(\hat{t}^{\mathrm{BH}}_m)=R_m^T(\hat{t}_m^{\mathrm{FDR}})$, but
can affect the risk\vspace*{1pt} $R^I_m(\cdot)$, that is,
$R^I_m(\hat{t}^{\mathrm{BH}}_m)\neq R^I_m(\hat {t}_m^{\mathrm{FDR}})$,
in general.\vspace*{1pt}

Finally, while relation (\ref{equ-FDRsupBonf}) uses $p$-values whereas
the algorithms defined in Section~\ref{secthresholdingproc} use
test statistics, it is easy to check that the resulting procedures are
the same.
%
%
\begin{remark}[(Adaptive FDR procedures under sparsity)]
To get a better FDR controlling procedure, one classical approach is to
modify (\ref{equ-FDRsupBonf}) by dividing $\alpha_m$ by a (more or
less explicit) estimator of $\pi_{0,m}$ and by possibly using a
step-up-down algorithm; see, for example, \cite
{TLD1998,Sar2002,BKY2006,Sar2008,BR2009,FDR2009,GBS2009}.
However, this method seems not helpful in our sparse setting because
$\pi_{0,m}$ is very close to $1$. As a result, we focus in this paper
only on the original (nonadaptive) version of FDR thresholding~(\ref
{equ-FDRsupBonf}).
\end{remark}

\section{Results in the general model}
\label{secgeneralresults}
This section presents relations of the form (\ref{equ-aim1}) and
(\ref{equ-aim2})
for the BFDR and FDR thresholds.
Our first main result deals with the BFDR threshold.
%
%
\begin{theorem}\label{main-th}
Assume (\ref{A}) and consider the BFDR threshold $t_m^\star$ at a
level $\alpha_m\in((1+f_m(0^+)/\tau_m)^{-1},\pi_{0,m})$
corresponding to a recovery parameter $q_m=\alpha_m^{-1}-1$. Consider
$q_m^{\mathrm{opt}}\geq1$ the optimal recovery parameter given by
(\ref{qmopt}).
Then the following holds:
\begin{longlist}[(ii)]
\item[(i)] if $\alpha_m\leq1/2$, we have for any $m\geq2$,
%
%
\begin{equation}
\label{mainbornsup-gen} R_m\bigl(t_m^\star
\bigr) - R_m\bigl(t_m^B\bigr)\leq
\pi_{1,m} \bigl\{\bigl(C_m/q_m -
C_m/q_m^{\mathrm{opt}} \bigr)\vee\gamma_m
\bigr\},
\end{equation}
where we let $\gamma_m=(C_m-F_m(\Psi_m^{-1}(q_m \tau_m)))_+$.

In particular, under (\ref{BP}), if $\alpha_m\rightarrow0$ and
$\gamma_m\rightarrow0$,
the BFDR threshold $t_m^\star$ is asymptotically optimal at rate $\rho
_m=\alpha_m+ \gamma_m$.
\item[(ii)] we have for any $m\geq2$,
%
%
\begin{equation}
\label{mainborninf} \frac{R_m(t_m^\star)}{R_m(t_m^B)} \geq\frac{\pi
_{1,m}}{R_m(t_m^B)} \bigl(1-
\bigl(1-q_m^{-1}\bigr)_+ F_m
\bigl(q_m^{-1} \tau_m^{-1} \bigr)
\bigr).
\end{equation}
In particular, under (\ref{BP}), if $R_m(t_m^B)\sim\pi_{1,m}
(1-C_m)$ and if $q_m$ is chosen such that
%
%
\begin{equation}
\label{mainborninfasymp} \liminf_m \biggl\{\frac{1-(1-q_m^{-1})_+ F_m(q_m^{-1}
\tau_m^{-1} )}{1-C_m} \biggr\}
>1,
\end{equation}
$t_m^\star$ is not asymptotically optimal.
\end{longlist}
\end{theorem}

Theorem~\ref{main-th} is proved in Section~\ref{secproofs}.
Theorem~\ref{main-th}(i) presents an upper-bound for the excess risk
when choosing $q_m$ instead of $q_m^{\mathrm{opt}}$ in BFDR thresholding.
First, both sides of (\ref{mainbornsup-gen}) are equal to zero when
$q_m=q_m^{\mathrm{opt}}$. Hence, this bound is\vadjust{\goodbreak} sharp in that case.
Second, assumption ``$\alpha_m\leq1/2$'' in Theorem~\ref{main-th}(i)
is only a technical detail that allows us to get $C_m/q_m$ instead of
$1/q_m$ in the right-hand side of (\ref{mainbornsup-gen}) (moreover,
it is quite natural; see the end of Section~\ref{secpFDRthresholding}).
Third, $\gamma_m$ has a simple interpretation as the difference
between the power of Bayes' thresholding and BFDR thresholding.
Fourth, bound (\ref{mainbornsup-gen}) induces the following trade-off
for choosing $\alpha_m$: on the one hand, $\alpha_m$ has to be chosen
small enough to make $C_m/q_m$ small; on the other hand, $\gamma_m$
increases as $\alpha_m$ decreases to zero.
Finally note that, in Theorem~\ref{main-th}(i), the second statement
is a consequence of the first one because $R_m(t_m^B)\geq\pi_{1,m} (1-C_m)$.
Theorem~\ref{main-th}(ii) states lower bounds
which are useful to identify regimes of $\alpha_m$ that do not lead to
an asymptotically optimal BFDR thresholding; see Corollary \ref
{cor-sub}(i) further on.

Our second main result deals with FDR thresholding.
%
%
\begin{theorem}\label{main-th2}
Let $\eps\in(0,1)$, assume (\ref{A}) and consider the FDR threshold
$\hat{t}^{\mathrm{FDR}}_m$ at level $\alpha_m>(1-\eps)^{-1}(\pi_{0,m}+\pi
_{1,m}f_m(0^+))^{-1}$.
Then the following holds: for any $m\geq2$,
%
%
\begin{eqnarray}
\label{mainbornsup2-gen} R_m\bigl(\hat{t}^{\mathrm{FDR}}_m
\bigr) - R_m\bigl(t_m^B\bigr) &\leq&
\pi_{1,m} \frac{\alpha_m}{1-\alpha_m} + m^{-1} \frac{\alpha_m}{(1-\alpha_m)^2}
\nonumber\\[-8pt]\\[-8pt]
&&{}+ \pi_{1,m} \bigl\{ \gamma'_{m} \wedge
\bigl(\gamma_{m}^\eps+ e^{-m\eps^2(\tau_m+1)^{-1} (C_m - \gamma_{m}^\eps
)/4 } \bigr) \bigr\}\nonumber
\end{eqnarray}
for $\gamma^\eps_{m}=(C_m-F_m(\Psi_m^{-1}(q_m^{\eps} \tau_m)))_+$
with $q_m^{\eps}= (\alpha_m\pi_{0,m}(1-\eps))^{-1}-1$ and $\gamma
'_{m}=(C_m-F_m(\alpha_m/m))_+$.
In particular, under (\ref{BP}) and assuming \mbox{$\alpha_m\rightarrow0$}:

\begin{longlist}[(ii)]
\item[(i)] if
$m/\tau_m \rightarrow+\infty$,
$\gamma_m^\eps\rightarrow0$ and additionally $\forall\kappa>0$,
$e^{-\kappa m/\tau_m}=o(\gamma^\eps_{m})$,
the FDR threshold $\hat{t}^{\mathrm{FDR}}_m$ is asymptotically optimal at rate
$\rho_m=\alpha_m+ \gamma_m^\eps$;\vspace*{2pt}
\item[(ii)] if $m/\tau_m \rightarrow\l\in(0,+\infty)$ with
$\gamma'_m\rightarrow0$,
the FDR threshold $\hat{t}^{\mathrm{FDR}}_m$ is asymptotically optimal at rate
$\rho_m=\alpha_m+ \gamma'_m$.
\end{longlist}
\end{theorem}

Theorem~\ref{main-th2} is proved in Section~\ref{secproofs}.
The proof mainly follows the methodology of~\cite{BCFG2010}, but is
more general and concise.
It relies on tools developed in \mbox{\cite
{FR2002,GW2002,FZ2006,FDR2009,RV2010,Roq2011}}.
The main argument for the proof is that the FDR threshold $\hat
{t}^{\mathrm{FDR}}_m(\alpha_m)$ is either well concentrated around the BFDR
threshold ${t}^\star_m(\alpha_m \pi_{0,m})$ (as illustrated in the
right panel of Figure~\ref{figexpoFDR}) or close to the Bonferroni
threshold $\alpha_m/m$. This argument was already used in~\cite{BCFG2010}.

Let us comment briefly on Theorem~\ref{main-th2}: first, as in the
BFDR case, choosing $\alpha_m$ such that the bound in (\ref
{mainbornsup2-gen}) is minimal involves a trade-off because $\gamma
_m^\eps$ and $\gamma'_m$ are quantities that increase when $\alpha_m$
decreases to zero.
Second, let us note that cases (i) and (ii) in Theorem~\ref{main-th2}
are intended to cover regimes where the FDR is close to BFDR
(moderately sparse) and where the FDR threshold is close to the
Bonferroni threshold (extremely sparse), respectively.
In particular, these two regimes cover the case where $\tau_m=m^\beta
$ with $\beta\in(0,1]$.
Finally, the bounds and convergence rates derived in Theorems \ref
{main-th} and~\ref{main-th2} strongly depend
on the nature of $F_m$. We derive a more explicit expression of the
latter in the next section, in the particular cases of location and
scale models coming from a Subbotin density.
%
%
\begin{remark}[(Conservative upper-bound for $\gamma_m$)]
\label{remupperboundgamma}
By the concavity of $F_m$, we have $q_m \tau_m=\Psi_m(t_m^\star)
\geq f_m(t_m^\star)$, which yields
%
%
\begin{equation}
\label{boundstmstar} \gamma_m\leq C_m- F_m
\bigl(f_m^{-1}(q_m \tau_m)\bigr)
\in[0,1).
\end{equation}
When $f_m^{-1}$ is easier to use than $\Psi_m^{-1}$, it is tempting to
use (\ref{boundstmstar}) to upper bound the excess risk in Theorems
\ref{main-th} and~\ref{main-th2}. However, this can inflate the
resulting upper-bound too much. This point is discussed in Section
10.4 in the supplemental article
\cite{NR2011supp} for the case of a Gaussian density (for which this
results in an additional $\log\log\tau_m$ factor in the
bound).
\end{remark}

\section{Application to location and scale models}\label
{secappl-locationandscale}

\subsection{The Bayes risk and optimal recovery parameter}
\label{secpourcommencer}

A preliminary task\vspace*{1pt} is to study the behavior of $t_m^B$, $R_m(t^B_m)$
and $q_m^{\mathrm{opt}}=C_m/(\tau_m t_m^B)$ both in location and
scale models.
While finite sample inequalities are given in Section~9.2 in the supplemental
article~\cite{NR2011supp}, we only
report in this subsection some resulting asymptotic relations for short.
Let us define the following rates, which will be useful throughout the paper:
%
%
\begin{eqnarray}
\label{rate-loc} r^{\mathrm{loc}}_m&=& \bigl(\zeta\log
\tau_m + \bigl|\overline{D} {}^{-1}(C_m)\bigr|^\zeta
\bigr)^{1-1/\zeta};
\\
\label{rate-scale} r^{\mathrm{sc}}_m&=& \zeta\log\tau_m
+\bigl(\overline{D} {}^{-1}(C_m/2)\bigr)^\zeta.
\end{eqnarray}
Under (\ref{Sp}), note that the rates $r^{\mathrm{loc}}_m$ (resp.,
$r^{\mathrm{sc}}_m$)
tend to infinity. Furthermore, by using Section 9.2 in
the supplemental article~\cite{NR2011supp}, we have $\mu_m=(r^{\mathrm{loc}}_m
)^{1/(\zeta-1)}- \overline{D}{}^{-1}(C_m)$ in the location model and
$\sigma_m\geq(r^{\mathrm{sc}}_m)^{1/\zeta} / (\overline{D}{}^{-1}(C_m/2))$
in the scale model.
%
%
\begin{proposition}\label{prop-Bayes}
Consider a $\zeta$-Subbotin density (\ref{equ-Subdensity}) with
$\zeta\geq1$ for a scale model and $\zeta> 1$ for a location model.
Let $(\tau_m,C_m)\in(1,\infty)\times(0,1)$ be the parameters of the model.
Let $r_m$ be equal to $r^{\mathrm{loc}}_m$ defined by (\ref
{rate-loc}) in the
location model or to $r^{\mathrm{sc}}_m$ defined by (\ref
{rate-scale}) in the
scale model.
Then, under (\ref{BP}) and (\ref{Sp}), we have $\mu_m \sim
r^{\mathrm{loc}}_m
\sim(\zeta\log\tau_m)^{1/\zeta}$ and $\sigma_m\sim(r^{\mathrm{sc}}_m
)^{1/\zeta}/(\overline{D}{}^{-1}(C_m/2)) \sim(\zeta\log\tau_m)^{1/\zeta
}/(\overline{D}{}^{-1}(C_m/2))$ and
%
%
\begin{eqnarray}
\label{equiRtmb} R_m\bigl(t_m^B\bigr)&\sim&
\pi_{1,m} (1-C_m),
\\
\label{majtmb} t^B_m&=&O \bigl( R_m
\bigl(t_m^B\bigr)/ r_m \bigr),
\\
\label{qmoptequiv} q_m^{\mathrm{opt}}&\sim&\cases{\displaystyle
\frac{C_m}{d(\overline
{D}{}^{-1}(C_m)) } (\zeta\log\tau_m)^{1-1/\zeta} &\quad(location),
\vspace*{2pt}\cr
\displaystyle\frac{C_m/2}{ \overline{D}{}^{-1}(C_m/2) d(\overline
{D}{}^{-1}(C_m/2)) } \zeta\log\tau_m &\quad(scale).}
\end{eqnarray}
\end{proposition}

From (\ref{equiRtmb}) and (\ref{majtmb}), by assuming (\ref{BP}) and
(\ref{Sp}), the probability of a type~I error ($\pi_{0,m}t_m^B$) is
always of smaller order than the probability of a type~II error ($\pi
_{1,m}(1-C_m)$). The latter had already been observed in \cite
{BCFG2010} in the particular case of a Gaussian scale model.
%
%
\begin{remark}\label{remrisque0}
From (\ref{equiRtmb}) and since the risk of null thresholding is
$R_m(0)=\pi_{1,m}$, a substantial improvement over the null threshold
can only be expected in the regime where $C_m\geq C_-$, where $C_-$ is
``far'' from $0$.
\end{remark}

\subsection{Finite sample oracle inequalities}\label{secoracleineq}

The following result can be derived from Theorem~\ref{main-th}(i) and
Theorem~\ref{main-th2}. It is proved in Section 9.3
in the supplemental article~\cite{NR2011supp}.
%
%
\begin{corollary}
\label{cor-oracle}
Consider a $\zeta$-Subbotin density (\ref{equ-Subdensity}) with
$\zeta> 1$ for a location model and $\zeta\geq1$ for a scale model,
and let $(\tau_m,C_m)\in(1,\infty)\times(0,1)$ be the parameters of
the model.
Let $r_m=r^{\mathrm{loc}}_m$ [defined by (\ref{rate-loc})] and
$K_m=d(0)$ in the
location model or $r_m=r^{\mathrm{sc}}_m$ [defined by (\ref
{rate-scale})] and
$K_m=2 \overline{D}{}^{-1}(C_m/2) d(\overline{D}{}^{-1}(C_m/2))$ in the
scale model.
Let $\alpha_m\in(0,1/2)$ and denote the corresponding recovery
parameter by $q_m=\alpha_m^{-1}-1$. Consider $q_m^{\mathrm{opt}}\geq
1$ the
optimal recovery parameter given by (\ref{qmopt}). Let $\nu\in
(0,1)$. Then:
\begin{longlist}[(ii)]
\item[(i)] The BFDR threshold $t_m^\star$ at level $\alpha_m$
defined by (\ref{thres-pFDR}) satisfies that for any $m\geq2$ such that
$r_m \geq\frac{K_m}{C_m(1-\nu)} (\log(q_m/q_m^{\mathrm{opt}})
-\log\nu)$,
%
%
\begin{eqnarray}
\label{mainbornsup} &&R_m\bigl(t_m^\star\bigr) -
R_m\bigl(t_m^B\bigr)
\nonumber\\[-8pt]\\[-8pt]
&&\qquad\leq\pi_{1,m} \biggl\{ \biggl(\frac{ C_m}{q_m}-
\frac{ C_m}{q_m^{\mathrm{opt}}} \biggr) \vee\biggl(K_m \frac{\log
(q_m/q_m^{\mathrm{opt}}) -\log\nu}{r_m} \biggr)
\biggr\}.\nonumber
\end{eqnarray}
\item[(ii)] Letting\vspace*{1pt} $\eps\in(0,1)$, $D_{1,m}= - \log(\nu\pi_{0,m}(1-\eps
))$ and $D_{2,m}= \log(\nu^{-1} \*C_m \tau_m^{-1}m)$,
the FDR threshold $\hat{t}^{\mathrm{FDR}}_m$ at level $\alpha_m $ defined by
(\ref{equ-FDRsupBonf}) satisfies that, for any $a\in\{1,2\}$, for any
$m\geq2$ such that $r_m \geq\frac{K_m}{C_m(1-\nu)} (\log(\alpha
_m^{-1}/q_m^{\mathrm{opt}}) + D_{a,m} )$,
%
%
\begin{eqnarray}
\label{mainbornsup2}\qquad R_m\bigl(\hat{t}^{\mathrm{FDR}}_m
\bigr) - R_m\bigl(t_m^B\bigr) &\leq&
\pi_{1,m} \biggl(\frac
{\alpha_m}{1-\alpha_m} + K_m\frac{( \log(\alpha_m^{-1}/q_m^{\mathrm{opt}})+
D_{a,m})_+}{r_m}
\biggr)
\nonumber\\[-8pt]\\[-8pt]
&&{}+ \frac{\alpha_m/m}{(1-\alpha_m)^2}+ \pi_{1,m} {\mathbf{1}\{a=1\}}
e^{-m(\tau_m+1)^{-1} \nu\eps^2 C_m /4 }.\nonumber
\end{eqnarray}
\end{longlist}
\end{corollary}

Corollary~\ref{cor-oracle}(ii) contains two distinct cases. The case
$a=1$ should be used when $m/\tau_m$ is large, because the remainder
term containing the exponential becomes small (whereas $D_{1,m}$ is
approximately constant). The case $a=2$ is intended to deal with the
regime where $m/\tau_m$ is not large, because $D_{2,m}$ is of the
order of a constant in that case.
The finite sample oracle inequalities (\ref{mainbornsup}) and (\ref
{mainbornsup2}) are useful to derive explicit\vadjust{\goodbreak} rates of convergence, as
we will see in the next section.
Let us also mention that an exact computation of the excess risk of
BFDR thresholding can be derived in the Laplace case; see Section
10.2 in the supplemental article~\cite{NR2011supp}.

\subsection{Asymptotic optimality with rates}\label{secrates}

In this section, we provide a sufficient condition on $\alpha_m$ such
that, under (\ref{BP}) and (\ref{Sp}), BFDR/FDR thresholding is
asymptotically optimal [according to (\ref{equ-aim2})], and we provide
an explicit rate $\rho_m$.
Furthermore, we establish that this condition is necessary for the
optimality of BFDR thresholding.
%
%
\begin{corollary}\label{cor-sub}
Take $\zeta>1$, $\gamma=1-\zeta^{-1}$ for the location case and
$\zeta\geq1$, $\gamma=1$ for the scale case.
Consider a $\zeta$-Subbotin density (\ref{equ-Subdensity}) in the
sparsity regime $\tau_m=m^{\beta}$, $0<\beta\leq1$ and under (\ref{BP}).
Then the following holds:
\begin{longlist}[(iii)]
\item[(i)] The BFDR threshold $t_m^\star$ is asymptotically optimal
if and only if
%
%
\begin{equation}
\label{cond-conv-sub} \alpha_m\rightarrow0 \quad\mbox{and}\quad\log
\alpha_m=o \bigl( (\log m)^\gamma\bigr),
\end{equation}
in which case it is asymptotically optimal at rate
$\rate=\alpha_m+\frac{ (\log(\alpha_m^{-1}/(\log m)^\gamma
))_+}{(\log m)^\gamma}.
$
\item[(ii)]
The FDR threshold $\hat{t}^{\mathrm{FDR}}_m$ at a level $\alpha_m$ satisfying
(\ref{cond-conv-sub}) is asymptotically optimal at rate $\rate=\alpha
_m+ \frac{(\log(\alpha_m^{-1}/(\log m)^\gamma))_+}{(\log
m)^\gamma}$.\vspace*{2pt}
\item[(iii)] Choosing $\alpha_m\propto1/(\log m)^\gamma$, BFDR and
FDR thresholding are both asymptotically optimal at rate $\rate
=1/(\log m)^\gamma$.
\end{longlist}
\end{corollary}

In the particular case of a Gaussian scale model $(\zeta=2)$,
Corollary~\ref{cor-sub} recovers Corollaries 4.2 and 5.1 of
\cite{BCFG2010}. Corollary~\ref{cor-sub} additionally provides a
rate, and encompasses the location case and other values of $\zeta$.
%
%
\begin{remark}[(Lower bound for the Laplace scale model)]
We can legitimately ask whether the rate $\rho_m=(\log m)^{-\gamma}$
can be improved. We show that this rate is the smallest that one can
obtain over a sparsity class $\beta\in[\beta_-,1]$ for some $\beta_-\in
(0,1)$, in the particular case of BFDR thresholding and in the
Laplace scale model; see Corollary 10.2 in the
supplemental article~\cite{NR2011supp}.
While the calculations become significantly more difficult in the other
models, we believe that the minimal rate for the relative excess risk
of the BFDR is still $(\log m)^{-\gamma}$ in a Subbotin location and
scale models.
Also, since the FDR can be seen as a stochastic variation around the
BFDR, we may conjecture that this rate is also minimal for FDR thresholding.
\end{remark}

\subsection{\texorpdfstring{Choosing $\alpha_m$}{Choosing alpha m}}\label{secchoicealpham}

Let us consider the sparsity regime $\tau_m=m^{\beta}$, $\beta\in(0,1)$.
Corollary~\ref{cor-sub} suggests to choose $\alpha_m$ such that
$\alpha_m\propto(\log m)^{-\gamma}$. This is in accordance with the
recommendation of~\cite{BCFG2010} in the Gaussian scale model; see
Remark~5.3 therein.
In this section, we propose an explicit choice of $\alpha_m$ from an
priori value
$(\beta_0,C_0)$ of the unknown parameter $(\beta,C_m)$.\vadjust{\goodbreak}

Let us choose a value $(\beta_0,C_0)$ a priori for $(\beta,C_m)$.
A natural choice for $\alpha_m$ is the value which would be optimal if
the\vspace*{1pt} parameters of the model were $(\beta,C_m)=(\beta_0,C_0)$. Namely,
by using (\ref{qmopt}) in Section~\ref{secpFDRthresholding}, we
choose $\alpha_m=\alpha_m^{\mathrm{opt}}(\beta_0,C_0)$, where
%
%
\begin{eqnarray}
\label{alphamopt0} \alpha_m^{\mathrm{opt}}(\beta_0,C_0)=
\bigl(1+q_m^{\mathrm
{opt}}(\beta_0,C_0)
\bigr)^{-1} \nonumber\\[-8pt]\\[-8pt]
&&\eqntext{\mbox{with $q_m^{\mathrm{opt}}(
\beta_0,C_0)=m^{-\beta_0} C_0 /
F_{m,0}^{-1}(C_0)$}}
\end{eqnarray}
by denoting $F_{m,0}$ the c.d.f. of the $p$-values following the
alternative for the model parameters $(\beta_0,C_0)$.
For instance:
\begin{itemize}[-]
\item[-] Gaussian location: $F_{m,0}^{-1}(C_0)= \overline{\Phi}
( \{\overline{\Phi}{}^{-1}(C_0)^2+2\beta_0 \log m \}^{1/2} )$;
\item[-] Gaussian scale: $F_{m,0}^{-1}(C_0)= 2\overline{\Phi}
(\overline{\Phi}{}^{-1}(C_0/2)x )$, where $x>1$ is the solution
of $2\beta_0 \log m + 2\log x= (\overline{\Phi}{}^{-1}(C_0/2))^2 (x^2-1)$;
\item[-] Laplace scale: $q_m^{\mathrm{opt}}(\beta_0,C_0)=y$, where
$y>1$ is the
solution of $\beta_0 \log m + \log y= (y-1)\log(1/C_0) $,
\end{itemize}
where $\overline{\Phi}(z)$ denotes $\P(Z\geq z)$ for $Z\sim\mathcal
{N}(0,1)$.

The above choice of $\alpha_m$ does depend on $(\beta_0,C_0)$, which
can be interpreted as a ``guess'' on the value of the unknown parameter
$(\beta,C_m)$. Hence, when no prior information on $(\beta,C_m)$ is
available from the data, the above choice of $\alpha_m$ can appear of
limited interest in practice. However, we would like to make the
following two points:
\begin{itemize}
\item asymptotically, choosing $\alpha_m=\alpha_m^{\mathrm
{opt}}(\beta_0,C_0)$
always yields an optimal\break (B)FDR thresholding [under (\ref{BP})], even
if $(\beta_0,C_0)\neq(\beta,C_m)$: by Proposition~\ref{prop-Bayes},
we get $\alpha_m^{\mathrm{opt}}(\beta_0,C_0)\propto(\log
m)^{-\gamma}$ and thus
the asymptotic optimality is a direct consequence of Corollary \ref
{cor-sub}(iii);
\item nonasymptotically, our numerical experiments suggest that
$\alpha_m=\break\alpha_m^{\mathrm{opt}}(\beta_0,C_0)$ performs fairly
well when
we have at hand an a priori on the location of the model parameters: if
$(\beta,C_m)$ is supposed to be in some specific (but possibly large)
region of the ``sparsity${}\times{}$power'' square, choosing any $(\beta
_0,C_0)$ in that region yields a thresholding procedure with a
reasonably small risk; see Sections~5 and 14
in the supplemental article~\cite{NR2011supp}.\looseness=-1
\end{itemize}

Finally, let us note that the choice $\alpha_m=\alpha_m^{\mathrm
{opt}}(\beta_0,C_0)$
is motivated by the analysis of the BFDR risk, not that of the FDR
risk. Hence, it might be possible to choose a better $\alpha_m$ for
FDR thresholding, especially for small values of $m$ for which BFDR and
FDR are different. Because obtaining such a refinement appeared quite
challenging, and as our proposed choice already performed well, we
decided not to investigate this question further.
%
%
\begin{remark}
\label{remDcomp}
By choosing $\alpha_m=\alpha_m^{\mathrm{opt}}(\beta_0,C_0)$ as in
(\ref
{alphamopt0}), we can legitimately ask how large the constants are in
the finite sample inequalities coming from Corollary~\ref{cor-oracle}
in standard cases. To simplify the problem, let us focus\vadjust{\goodbreak} on the BFDR
threshold and
consider a $\zeta$-Subbotin location model with $\zeta> 1$. Taking
$\tau_m=m^{\beta}$, the parameters of the model are $(\beta,C_m)\in
(0,1]\times(0,1)$. Assume that the parameter sequence $(C_m)_m$
satisfies (\ref{BP}) for some $0<C_-\leq C_+<1$.
Then Corollary 9.4 in the supplemental article \cite
{NR2011supp} provides explicit constants $D=D(\beta,C_-,C_+, \beta_0,
C_0,\nu)$ and $M=M(\beta,C_-,C_+, \beta_0, C_0,\nu)$ such that
the following inequality holds:
%
%
\begin{equation}
\label{oracleineqwithconstants} \qquad\bigl(R_m\bigl(t_m^\star
\bigr) - R_m\bigl(t_m^B\bigr) \bigr)/
R_m\bigl(t_m^B\bigr)\leq D / (\log
m)^{1-1/\zeta}\qquad\mbox{for any $m\geq M$.}
\end{equation}
As an illustration, in the Gaussian case ($\zeta=2$), for $\beta
=0.7$, $C_-=0.5$, $C_+=0.7$, $\beta_0=C_0=0.5$ and $\nu=0.25$, we
have $M\simeq61.6$ and $D\simeq2.66$. As expected, these constants
are over-estimated: for instance, by taking $m=1000$, the left-hand
side of (\ref{oracleineqwithconstants}) is smaller than $0.1$ (see
Figure~\ref{figgaussian-location-heatmap-C0=05} in the next section)
while the right-hand side of (\ref{oracleineqwithconstants}) is
$D/\sqrt{\log(1000)}\simeq1.01$.
Finally,
we can check that $D$ becomes large when $\beta$ is close to $0$ or
$C_+$ is close to $1$.
These configurations correspond to the cases where the data are almost
nonsparse and where the Bayes rule can have almost full power,
respectively. They can be seen as limit cases for our methodology.
\end{remark}
%
%
\begin{remark}
By using Proposition~\ref{prop-Bayes}, as $m \to+\infty$,
$\alpha_m^{\mathrm{opt}}(\beta_0,C_0)\sim\alpha_m^{\infty}(\beta
_0,C_0)$, for an
equivalent $\alpha_m^{\infty}(\beta_0,C_0)$ having a very simple form;
see Section 10.1 in the supplemental article \cite
{NR2011supp}. Therefore, we could use $\alpha_m^{\infty}(\beta_0,C_0)$
instead of $\alpha_m^{\mathrm{opt}}(\beta_0,C_0)$.
Numerical comparisons between the (B)FDR risk obtained according to
$\alpha_m^{\mathrm{opt}}(\beta_0,C_0)$ and $\alpha_m^{\infty
}(\beta_0,C_0)$ are
provided in Section 14 in the supplemental article \cite
{NR2011supp}. While $\alpha_m^{\infty}(\beta_0,C_0)$ qualitatively
leads to
the same results when $m$ is large (say, $m\geq1000$), the use of
$\alpha_m^{\mathrm{opt}}(\beta_0,C_0)$ is more accurate for a small $m$.
\end{remark}

\section{Numerical experiments}\label{secnumexp}
In order to complement the convergence results stated above, it is of
interest to study the behavior of FDR and BFDR thresholding for a small
or moderate $m$ in numerical experiments. These experiments have been
performed for the inductive risk $R_m(\cdot)=R^I_m(\cdot)$ defined by
(\ref{defrisk}).

\subsection{Exact formula for the FDR risk}
The BFDR threshold $t_m^{\star}$ can be approximated numerically,
which allows us to compute $R_m(t_m^\star)$. Computing $R_m(\hat
{t}^{\mathrm{FDR}}_m)$ is more complicated because the FDR threshold $\hat
{t}^{\mathrm{FDR}}_m$ is not deterministic. However, we can avoid performing
cumbersome and somewhat imprecise simulations to compute
%
%
\begin{figure}

\includegraphics{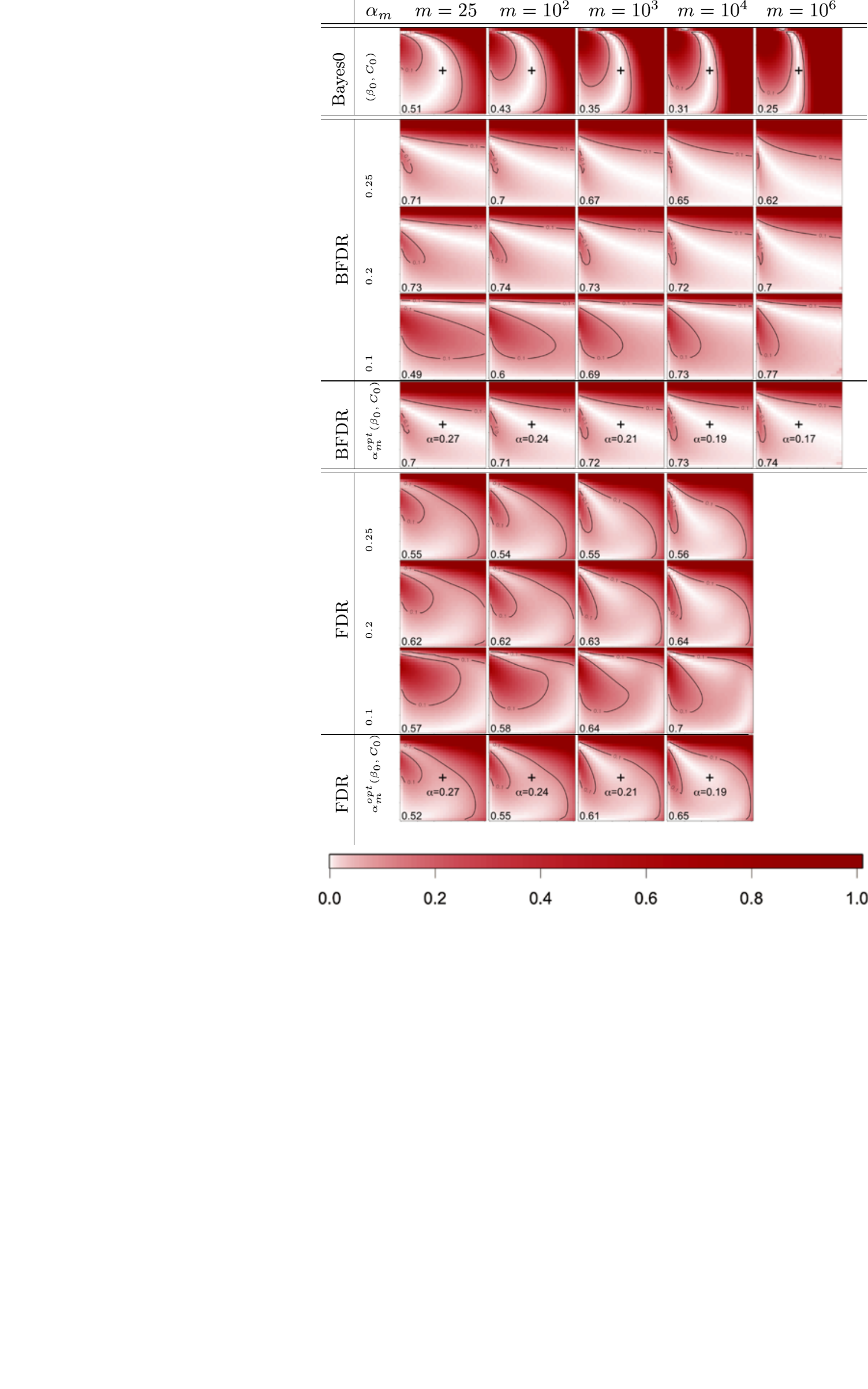}

\caption{Adaptation to sparsity by (B)FDR thresholding in the Gaussian
location model relative excess risks $\mathcal{E}_m$ for various
thresholding procedures (rows) and different values of $m$ (columns).
In each panel, the corresponding risk is plotted as a function of
$\beta\in[0,1]$ (horizontal axis) and $C_m \in[0, 1]$ (vertical
axis). Colors range from white (low risk) to dark red (high risk), as
indicated by the color bar at the bottom.
Black lines represent the level set $\mathcal{E}_m=0.1$. The point
$(\beta,C_m)=(\beta_0,C_0)$ is marked by ``$+$.'' We chose $\beta
_0=1/2$ and $C_0=1/2$. See main text for details.}
\label{figgaussian-location-heatmap-C0=05}
\end{figure}
$R_m(\hat
{t}^{\mathrm{FDR}}_m)$ by using the approach proposed in~\cite{FR2002} and
\cite{RV2010}. Using this methodology, the full distribution of $\hat
{t}^{\mathrm{FDR}}_m$ may be written as a function of the joint c.d.f. of the
order statistics of i.i.d. uniform variables.
Let for any $k\geq0$ and for any $(t_1,\ldots,t_k)\in[0,1]^k$,
$
\Psi_k(t_1,\ldots,t_k)= \P(U_{(1)}\leq t_1,\ldots,U_{(k)}\leq t_k),
$
where $(U_i)_{1\leq i \leq k}$ is a sequence of i.i.d. uniform
variables on $(0,1)$ and with the convention $\Psi_0(\cdot)=1$. The
$\Psi_k$'s can be evaluated, for example, by using Steck's recursion;\vadjust{\goodbreak}
see~\cite{SW1986}, pages 366--369.
Then, relation (10) in~\cite{RV2010} entails
%
%
\begin{eqnarray}
\label{exactFDR} R_m\bigl(\hat{t}^{\mathrm{FDR}}_m\bigr)
&=& \sum_{k=0}^m \pmatrix{m
\cr
k}
R_m \biggl(\frac
{\alpha(k \vee1)}{m} \biggr) G_m(\alpha
k/m)^k
\nonumber\\[-9pt]\\[-10pt]
&&\hspace*{14pt}{}\times\Psi_{m-k} \bigl(1-G_m(\alpha m/m),
\ldots,1-G_m\bigl(\alpha(k+1)/m\bigr) \bigr),\nonumber
\end{eqnarray}
where $G_m(t)=\pi_{0,m} t + \pi_{1,m}F_m(t)$.
For reasonably large $m$ ($m\leq10\mbox{,}000$ in what follows), expression
(\ref{exactFDR}) can be used for computing the \textit{exact} risk of
FDR thresholding $\hat{t}^{\mathrm{FDR}}_m$ in our
experiment.\vspace*{-2pt}

\subsection{Adaptation to unknown sparsity}\label{secprior}

We quantify the quality of a thresholding procedure using the relative
excess risk
\[
\mathcal{E}_m(\thr)=\bigl(R_m(\thr)-R_m
\bigl(t^B_m\bigr)\bigr)/R_m
\bigl(t^B_m\bigr).
\]
The closer the relative excess risk $\mathcal{E}_m(\hat{t}_m)$ is to
$0$, the better the corresponding classification procedure is.

Figure~\ref{figgaussian-location-heatmap-C0=05} compares relative
excess risks of different procedures in the Gaussian location model
(results for the Gaussian scale and the Laplace scale models are
qualitatively similar, see Figures 5 and 6 in the supplemental article
\cite{NR2011supp}).
Each row of plots corresponds to a particular procedure, and each
column to a particular value of $m\in\{25, 10^2, 10^3, 10^4, 10^5,
10^6\}$.
The first row corresponds to the Bayes procedure defined by (\ref
{thres-Bayes}), where the model parameters are taken as $(\beta,
C_m)=(\beta_0,C_0)$. It is denoted by \textit{Bayes0}. Next, we
consider BFDR (rows 2 to 5) and FDR (rows 6 to 9) thresholding at level
$\alpha_m$, for $\alpha_m \in\{0.1, 0.2, 0.25 \}$ (independent of
$m$) and for the choice $\alpha_m=\alpha_m^{\mathrm{opt}}(\beta_0,
C_0)$ defined
in Section~\ref{secchoicealpham}.
For each procedure and each value of $m$, the behavior of the relative
excess risk is studied as the (unknown) true model parameters $(\beta,
C_m)$ vary in $[0,1] \times[0,1]$, and we arbitrarily choose $\beta_0$
and $C_0$ as the midpoints of the corresponding intervals, that is,
$(\beta_0,C_0)=(1/2,1/2)$ (similar results are obtained for other
values of $(\beta_0,C_0)$; see Figures 8,
9 and 10 in the supplemental article
\cite{NR2011supp}). Colors reflect the value of the relative excess
risk. They range from white [$R_m=R_m(t^B_m)$] to dark red [$R_m\geq
2R_m(t^B_m)$].
Black lines represent the level set $\mathcal{E}_m=0.1$, that is, they
delineate a region of the $(\beta, C_m)$ plane in which the excess
risk of the procedure under study is ten times less than the Bayes risk.
The number at the bottom left of each plot gives the fraction of
configurations $(\beta, C_m)$ for which $\mathcal{E}_m \leq0.1$.
This evaluates the quality of a procedure uniformly across all the
$(\beta, C_m)$ values.

For $m=10^6$, we did not undertake exact FDR risk calculations: they
were too computationally intensive, as the complexity of the
calculation of function $\Psi_k$ used in (\ref{exactFDR}) is
quadratic in $m$. However, FDR risk is expected to be well approximated
by BFDR risk for such a large value of $m$, as confirmed by the fact
that FDR and BFDR plots at a given level $\alpha$ are increasingly
similar as $m$ increases.

Bayes0 performs well when the sparsity parameter $\beta$ is correctly
specified, and its performance is fairly robust\vadjust{\goodbreak} to $C_m$. However, it
performs poorly when $\beta$ is misspecified, and increasingly so as
$m$ increases.
The results are markedly different for the other thresholding methods.
BFDR thresholding and FDR thresholding are less adaptive to $C_m$ than
Bayes0, but much more adaptive to the sparsity parameter $\beta$, as
illustrated by the fact that the configurations with low relative
excess risk span the whole range of $\beta$.

For $\alpha_m=\alpha_m^{\mathrm{opt}}(\beta_0,C_0)$, the fraction of
configurations $(\beta, C_m)$ for which $\mathcal{E}_m \leq0.1$
increases as $m$ increases.
This illustrates the asymptotic optimality of (B)FDR thresholding, as
stated in Corollary~\ref{cor-sub}(iii), because $\alpha_m^{\mathrm
{opt}}(\beta_0,C_0)\propto(\log m)^{-1/2}$.
Additionally, observe that the $(\beta,C_m)$-region around $(\beta
_0,C_0)$ contains only very small values of $\mathcal{E}_m$, even for
moderate $m$.
This suggests that, nonasymptotically, $\alpha_m^{\mathrm
{opt}}(\beta_0,C_0)$ is
a reasonable choice for $\alpha_m$, when we know a priori that the
parameters lie in some specific region of the $(\beta,C_m)$-square.

Next, let us consider the case of (B)FDR thresholding using a fixed
value of $\alpha_m=\alpha$. While our theoretical results show that
choosing $\alpha_m$ fixed with $m$ (and in particular not tending to
zero) is always asymptotically sub-optimal, the results shown by
Figure~\ref{figgaussian-location-heatmap-C0=05} are less clear-cut.
An explanation is that $(\log m)^{-1/2}$ decreases only slowly to zero
[e.g., $\alpha_m^{\mathrm{opt}}(\beta_0,C_0)\simeq0.17$ for
$m=10^6$], hence the
asymptotic is quite ``far'' and not fully attained by our experiments.

Hence, from a more practical point of view, in a classical situation
where $m$ does not exceed, say, $10^6$, a practitioner willing to use
the (B)FDR can consider two different approaches to calibrate $\alpha
_m$: the first one is to take some arbitrary value, for example,
$0.05$, $0.1$ or $0.2$. The overall excess risk might be small, but the
location of the region of smallest excess risk (pictured in white in
our figures) is unknown, and depends strongly on $\alpha$ and $m$ (and
even $\zeta$). In contrast, the second method $\alpha_m^{\mathrm
{opt}}(\beta_0,C_0)$ ``stabilizes'' the region of the $(\beta
,C_m)$-square where
the (B)FDR has good performance across all the values of $m$ (and
$\zeta$).
Thus, while the first method has a clear interpretation in terms of
FDR, the second approach is more interpretable w.r.t. the sparsity and
power parameters and is recommended when these parameters are felt to
correctly parametrize the model.

Finally note that, when considering the weighted mis-classification
risk (as formally defined in (\ref{defw-risk}) and studied in
Section 11 in the supplemental article \cite
{NR2011supp}), there exists a particular choice of the weight (as a
function of $m$) such that the optimal (B)FDR level $\alpha_m^{\mathrm
{opt}}(\beta_0,C_0)$ does not depend on $m$, making (B)FDR thresholding
with fixed
values of $\alpha_m$ asymptotically optimal, as noted by \cite
{BCFG2010}. This point is discussed in Section~\ref{secweighting}.

\section{Discussion}

\subsection{Asymptotic minimaxity over a sparsity class}
\label{secdiscussrate}\label{secrelationpreviouswork}

Let us consider the sparsity range $\tau_m=m^{\beta}$, with $\beta_-\leq
\beta\leq1$, for some given $\beta_-\in(0,1)$. Assume (\ref
{BP}) with $C_-$ and $C_+$ defined therein.\vadjust{\goodbreak}
Denote the set $[\beta_-,1]\times[C_-,C_+]$ by $\Theta$ for short.
The minimax risk is defined by
\[
R_m^{\star}=\inf_{\hat{t}_m} \Bigl\{ \sup_{(\beta,C_m)\in\Theta
}
\bigl\{R_m(\hat{t}_{m}) \bigr\} \Bigr\},
\]
where the infimum is taken over the set of thresholds that can be
written as measurable functions of the $p$-values. Obviously,
$R_m^{\star} \geq\sup_{(\beta,C_m)\in\Theta}\{R_m({t}_{m}^B)\}$,
where ${t}_{m}^B$ is the Bayes threshold.
Hence, by taking the supremum w.r.t. $(\beta,C_m)$ in our excess risk
inequalities, we are able to derive minimax results.
However, this requires a precise formulation of (\ref{equ-aim2}) where
the dependence in $\beta$ of the constant $D$
is explicit.
For simplicity, let us consider the Laplace scale model. By using
(69) and (74) in the
supplemental article~\cite{NR2011supp}, and by taking $\alpha_m
\propto(\log m)^{-1}$, we can derive that there exists a constant
$D'>0$ (independent of $\beta_-$, $C_-$ and $C_+$) such that for a
large $m$,
%
%
\begin{eqnarray}
\label{equ-minimax?} \sup_{(\beta,C_m)\in\Theta} \bigl\{R_m\bigl(
\thr^{\mathrm{FDR}}\bigr) \bigr\} &\leq& \sup_{(\beta,C_m)\in\Theta} \bigl
\{R_m\bigl({t}_{m}^B\bigr) \bigr\} \biggl( 1+
\frac{-\log(\beta_-/2)}{\beta_-(1-C_+)}\frac{D'}{\log m} \biggr)
\nonumber\\[-8pt]\\[-8pt]
&\leq& R_m^{\star} \biggl( 1+\frac{-\log(\beta_-/2)}{\beta_-(1-C_+)}
\frac{D'}{\log m} \biggr).\nonumber
\end{eqnarray}
This entails that $\thr^{\mathrm{FDR}}$ is asymptotically minimax, that is,
\[
\sup_{(\beta,C_m)\in\Theta} \bigl\{R_m\bigl(\thr^{\mathrm{FDR}}\bigr) \bigr
\} \sim R_m^{\star}.
\]
This property can be seen as an analogue to the asymptotically
minimaxity stated in Theorem 1.1 in~\cite{ABDJ2006} and Theorem 1.3 in
\cite{DJ2006}, in an estimation context.

Finally, regarding (\ref{equ-minimax?}), an interesting avenue for
future research would be to establish whether there are asymptotically
minimax rules $\thr$ such that $\sup_{(\beta,C_m)\in\Theta} \{
R_m(\thr) \} = R_m^{\star} (1+o(\rho_m))$ for a rate $\rho_m$ smaller
than $(\log m)^{-1}$.

\subsection{Extension to weighted mis-classification risk}
\label{secweighting}

In our sparse setting, where we assume that there are many more labels
``$0$'' than labels ``$1$,'' one could consider that mis-classifying a
``$0$'' is less important than mis-classifying a ``$1$.'' This suggests
to consider the following weighted risk:
%
%
\begin{equation}
\label{defw-risk} R_{m,\lambda_m}(\hat{t}_m) = \E\bigl(
\pi_{0,m}\hat{t}_m + \lambda_m \pi_{1,m}
\bigl(1-F_m(\hat{t}_m)\bigr) \bigr)
\end{equation}
for a known factor $\lambda_m\in(1,\tau_m)$. This weighted risk was
extensively used in~\cite{BCFG2010}.
In Section 11 in the supplemental article \cite
{NR2011supp}, we show that all our results can be adapted to this risk.
Essentially, when considering $R_{m,\lambda_m}$ instead of $R_{m}$,
our results hold after replacing $\tau_m$ by $\tau_m/\lambda_m$ and
$q_m$ by $q_m\lambda_m$.

As an illustration, let us consider here the case of a $\zeta
$-Subbotin density, $\tau_m=m^\beta$, $\beta\in(0,1]$, $\log
\lambda_m=o((\log m)^\gamma)$, where $\gamma=1-\zeta^{-1}$ and
$\gamma=1$ for the location and scale cases,\vadjust{\goodbreak} respectively.
As displayed in Table 4 in the supplemental
article~\cite{NR2011supp}, under the (corresponding) assumptions
(\ref{BP}) and (\ref{Sp}), we show that a sufficient condition for
FDR thresholding to be asymptotically optimal for the risk
$R_{m,\lambda_m}$ is to take $q_m^{-1}=O(1)$, $q_m\lambda_m\rightarrow
\infty$ and $\log q_m=o ((\log m)^\gamma)$.
This recovers Theorem 5.3 of~\cite{BCFG2010} when applied to the
particular case of a Gaussian scale model (for which $\gamma=1$).
Furthermore, we show that taking $q_m \propto q_m^{\mathrm{opt}}$,
that is, $q_m
\propto\lambda_m^{-1} (\log m)^{\gamma} $, leads to the optimality
rate $\rho_m=(\log m)^{-\gamma}$ for the relative excess risk based
on $R_{m,\lambda_m}$.
While the order of $q_m^{\mathrm{opt}}$ is not modified when $\lambda
_m\propto
1$, it may be substantially different when $\lambda_m\rightarrow
\infty$. Typically, $\lambda_m \propto(\log m)^{\gamma}$ leads to
$q_m^{\mathrm{opt}}\propto1$. Hence, when considering $R_{m,\lambda
_m}$ instead
of $R_{m}$, the value of $\lambda_m$ should be carefully taken into
account when choosing $\alpha_m$ to obtain a small excess risk.

Conversely, our result states that FDR thresholding with a
pre-specified value of $\alpha_m=\alpha$ (say, $\alpha=0.05$), is
optimal over the range of weighted mis-classification risks using a
$\lambda_m$ satisfying $\lambda_m\rightarrow\infty$ and $\log
\lambda_m=o ((\log m)^\gamma)$, and that choosing
$\lambda_m \propto(\log m)^{\gamma}$ leads to the optimality rate
$\rho_m=(\log m)^{-\gamma}$.

\section{\texorpdfstring{Proofs of Theorems \protect\ref{main-th} and \protect\ref{main-th2}}
{Proofs of Theorems 3.1 and 3.2}}\label{secproofs}

The proofs are first established for the misclassification risk
$R_m={R}^I_m$ defined by (\ref{defrisk}). The case of the
misclassification risk ${R}^T_m$, defined by (\ref{defrisk-bis}) is
examined in Section~\ref{secotherrisk}.

\subsection{Relations for BFDR}
Let us first state the following result.
%
%
\begin{proposition}\label{sharper-oipFDR}
Consider the setting and the notation of Theorem~\ref{main-th}. Then
we have
for any $m\geq2$,
%
%
\begin{eqnarray}
\label{oracleequality} R_m\bigl(t_m^\star\bigr)
- R_m\bigl(t_m^B\bigr) &=& \pi_{1,m}
C_m/q_m- \pi_{0,m} t_m^B\nonumber\\[-8pt]\\[-8pt]
&&{}+
\pi_{1,m} \bigl(1-q_{m}^{-1}\bigr)
\bigl(C_m-F_m\bigl(t_m^\star\bigr)
\bigr).\nonumber
\end{eqnarray}
Furthermore,
if $\alpha_m\leq1/2$, we have for any $m\geq2$,
%
%
\begin{eqnarray}
\label{mainbornsup-improv1} R_m\bigl(t_m^\star
\bigr)- R_m\bigl(t_m^B\bigr)  &\leq&
\pi_{1,m} C_m/q_m- \pi_{0,m}
t_m^B+ \pi_{1,m} \bigl(1-q_{m}^{-1}
\bigr)\gamma_m;
\\
\label{mainbornsup-improv2} R_m\bigl(t_m^\star
\bigr)- R_m\bigl(t_m^B\bigr) &\leq&
\pi_{1,m} \bigl(C_m/q_m- \tau_m
t_m^B\bigr)\vee\gamma_m.
\end{eqnarray}
\end{proposition}
\begin{pf}
To prove (\ref{oracleequality}), we use $F_m(t_m^\star)= t_m^\star
q_m\tau_m $ and $\tau_m=\pi_{0,m}/\pi_{1,m}$, to write
%
%
\begin{eqnarray}
\label{equ-interm-plus} && R_m\bigl(t_m^\star
\bigr)- R_m\bigl(t_m^B\bigr)
\nonumber
\\
&&\qquad= \pi_{0,m} t_m^\star-\pi_{0,m}
t_m^B + \pi_{1,m} \bigl(C_m-
F_m\bigl(t_m^\star\bigr)\bigr)
\\
&&\qquad=\pi_{1,m} F_m\bigl(t_m^\star
\bigr)/q_m -\pi_{0,m} t_m^B +
\pi_{1,m} \bigl(C_m- F_m\bigl(t_m^\star
\bigr)\bigr).
\nonumber
\end{eqnarray}
Expression (\ref{mainbornsup-improv1}) is an easy consequence of
(\ref{oracleequality}).
Finally, (\ref{equ-interm-plus}) and (\ref{oracleequality}) entail
\[
R_m\bigl(t_m^\star\bigr)- R_m
\bigl(t_m^B\bigr)\leq\cases{ \pi_{1,m}
C_m /q_m - \pi_{0,m} t_m^B,
&\quad if $t_m^B\leq t_m^\star$,
\vspace*{2pt}\cr
\pi_{1,m} \bigl(C_m-F_m\bigl(
\Psi_m^{-1}(q_m \tau_m)\bigr)\bigr),
&\quad if $t_m^B\geq t_m^\star$,}
\]
which yields (\ref{mainbornsup-improv2}).\vadjust{\goodbreak}
\end{pf}

\subsection{\texorpdfstring{Proof of Theorem \protect\ref{main-th}}{Proof of Theorem 3.1}}\label{secproofs-pFDR}

Theorem~\ref{main-th}(i) follows from (\ref{mainbornsup-improv2})
because $\pi_{0,m}\* t_m^B=\pi_{1,m} C_m/q_m^{\mathrm{opt}}$ by definition.
Let us now prove (ii). First note that
%
%
\begin{equation}
\label{riskFm} R_m\bigl(t_m^\star\bigr)=
\pi_{1,m}- \pi_{1,m} F_m\bigl(t_m^\star
\bigr) \bigl(1-q_m^{-1}\bigr).
\end{equation}
Using (\ref{riskFm}) and the upper bound $ t_m^\star=F_m(t_m^\star)
(q_m\tau_m)^{-1} \leq(q_m\tau_m)^{-1} $, we obtain
$R_m(t_m^\star)
\geq\pi_{1,m} (1- (1-q_m^{-1})_+ F_m(t_m^\star))
\geq\pi_{1,m} (1- (1-q_m^{-1})_+ F_m(q_m^{-1} \tau_m^{-1}))
$.
This entails (\ref{mainborninf}) and (\ref{mainborninfasymp}).

\subsection{\texorpdfstring{Proof of Theorem \protect\ref{main-th2}}{Proof of Theorem 3.2}}\label{secproofs-FDR}

Write $\hat{t}_m$ instead of $\hat{t}^{\mathrm{FDR}}_m$ for short.
To establish (\ref{mainbornsup2-gen}), let us first write the risk of
FDR thresholding as
$R_m(\hat{t}_m) = T_{1,m} + T_{2,m}$,
with
$T_{1,m}= \pi_{0,m} \E( \hat{t}_m) $ and
$T_{2,m}= \pi_{1,m} (1- \E(F_m(\hat{t}_m)) )$.
In the sequel, $T_{1,m}$ and $T_{2,m}$ are examined separately.

\subsubsection{Bounding $T_{1,m}$}

The next result is a variation of Lemmas 7.1 and~7.2 in~\cite{BCFG2010}.
%
%
\begin{proposition}\label{prop-BCFG}
The following bound holds:
%
%
\begin{equation}
\label{erreurtype1} T_{1,m}\leq\pi_{1,m} \frac{\alpha_m}{1-\alpha_m} +
m^{-1} \frac
{\alpha_m}{(1-\alpha_m)^2} .
\end{equation}
\end{proposition}
\begin{pf}
To prove Proposition~\ref{prop-BCFG}, we follow the proof of Lemma 7.1
in~\cite{BCFG2010} with slight simplifications.
Recall that we have by definition $\hat{t}_m=\hat{t}_m^{\mathrm{BH}} \vee
(\alpha_m /m)$.
Hence, we have $\E(\hat{t}_m | H)\leq\alpha_m/m + \E(\hat
{t}_m^{\mathrm{BH}} | H)$.
By integrating w.r.t. the label vector $H$, it is thus sufficient to prove
%
%
\begin{equation}
\label{toproveinprop-BCFG} E\bigl(\hat{t}_m^{\mathrm{BH}} | H\bigr)
\leq\pi_{1,m} \frac{\alpha_m}{1-\alpha_m} + m^{-1} \frac{\alpha
_m^2}{(1-\alpha_m)^2}.
\end{equation}
Let $m_1(H)=\sum_{i=1}^m H_i$ and $m_0(H)=m-m_1(H)$.
By exchangeability of $(p_i,H_i)_i$, we can assume without loss of
generality that the $p$-values corresponding to a label $H_i=0$ are
$p_{1},\ldots,p_{m_0(H)}$ for simplicity.
Let us denote $\hat{t}_{m,0}$ the thresholding $\thr^{\mathrm{BH}}$ defined by
(\ref{equFDR}), applied to the $p$-value family $p_i,1\leq i\leq m$,
in which each of the $p$-value $p_{m_0(H)+1},\ldots,p_{m}$ has been
replaced by $0$. Classically, we have
\[
\hat{t}_{m,0}=\alpha_m \bigl(m_1(H)+
\hat{k}_{m,0}\bigr)/m,
\]
where $\hat{k}_{m,0}= \max\{k\in\{0,1,\ldots,m_0(H)\}\dvtx
q_{(k)}\leq\alpha_m (m_1(H)+k)/m\}$,
where $(q_1,\ldots,q_{m_0(H)})=(p_{1},\ldots,p_{m_0(H)})$ is the set of
$p$-values corresponding to zero labels; see, for example, Lemma 7.1 in
\cite{RV2010}.
Since $\thr^{\mathrm{BH}}$ is nonincreasing in each $p$-value, setting some
$p$-values equal to $0$ can only increase $\thr^{\mathrm{BH}}$. This entails
%
%
\begin{equation}
\label{equ-interm-T1m} \E\bigl( \thr^{\mathrm{BH}} \cond H\bigr) \leq\E(
\hat{t}_{m,0} \cond H) = \alpha_m \bigl(m_1(H)+
\E(\hat{k}_{m,0} \cond H)\bigr)/m.
\end{equation}
Next, we use Lemma 4.2 in~\cite{FR2002} [by taking ``$n=m_0(H)$,
$\beta=\alpha_m$, $\tau=\alpha_m/m$'' with their notation], to
derive that for any $H\in\{0,1\}^m$,
%
%
\begin{eqnarray}
\label{equsansnom}\qquad
\E(\hat{k}_{m,0}\cond H) &=& \alpha_m
\frac{m_0(H)}{m} \sum_{i=0}^{m_0(H)-1}
\pmatrix{m_0(H)-1
\cr
i} \bigl(m_1(H)+i+1\bigr) i!
\biggl(\frac
{\alpha_m}{m} \biggr)^i
\nonumber
\\
&\leq&\alpha_m \sum_{i\geq0}
\bigl(m_1(H)+i+1\bigr) \alpha_m^i
\\
&=&\alpha_m \bigl(m_1(H)/(1-\alpha_m)
+1/(1-\alpha_m)^2\bigr).\nonumber
\end{eqnarray}
Bound (\ref{toproveinprop-BCFG}) thus follows from (\ref{equ-interm-T1m}).
\end{pf}

\subsubsection{Bounding $T_{2,m}$}

Let us consider $t_m^{\eps}$ the BFDR threshold associated to level
$\alpha_m\pi_{0,m}(1-\varepsilon)$.
Note that by definition of $t_m^{\eps}$ we have $\pi_{0,m}(1-\eps
)G_m(t_m^{\eps})=t_m^{\eps}/\alpha_m$.
Here, we state the following inequalities, which, combined with
Proposition~\ref{prop-BCFG} establishes Theorem~\ref{main-th2}:
%
%
\begin{eqnarray}\qquad
\label{erreurtype2-bonf} T_{2,m}&\leq&\pi_{1,m}
\bigl(1-F_m(\alpha_m/m)\bigr) ;
\\
\label{erreurtype2} T_{2,m}&\leq&\pi_{1,m}
\bigl(1-F_m\bigl(t_m^{\eps}\bigr)\bigr) +
\pi_{1,m} \exp\bigl\{-m (\tau_m+1)^{-1}
\bigl(C_m - \gamma_{m}^\eps\bigr)
\eps^2/4 \bigr\}.
\end{eqnarray}
First, (\ref{erreurtype2-bonf}) is an easy consequence of $\thr\geq
\alpha_m/m$. Second, expression (\ref{erreurtype2}) derives from
(\ref{conc-borneinf}) of Lemma~\ref{lemconc} because
\begin{eqnarray*}
\E\bigl(1-F_m(\thr)\bigr) &=& \E\bigl(\bigl(1-F_m(\thr)
\bigr){\mathbf{1}\bigl\{\thr<t_m^{\eps}\bigr\} }\bigr) + \E
\bigl(\bigl(1-F_m(\thr)\bigr){\mathbf{1}\bigl\{\thr\geq
t_m^{\eps}\bigr\}}\bigr)
\\
&\leq&\P\bigl(\thr^{\mathrm{BH}} < t_m^{\eps}\bigr) +
1-F_m\bigl(t_m^{\eps}\bigr)
\end{eqnarray*}
(by using $\thr\!\geq\!\thr^{\mathrm{BH}}$) and because \mbox{$G_m(t_m^{\eps})\!\geq\!\pi
_{1,m} F_m(t_m^{\eps} )\!\geq\!(\tau_m\!+\!1)^{-1}(C_m\!-\!\gamma_{m}^\eps)$}.

%
\begin{lemma}\label{lemconc}
The following bound holds:
%
%
\begin{equation}
\label{conc-borneinf} \P\bigl(\thr^{\mathrm{BH}} < t_m^{\eps}
\bigr) \leq\exp\bigl\{-mG_m\bigl(t_m^{\eps}\bigr)
\eps^2/4 \bigr\}.
\end{equation}
\end{lemma}

We prove Lemma~\ref{lemconc} by using
a variation of the method described in the proof of Theorem 1 in \cite
{GW2002} (we use Bennett's inequality instead of Hoeffding's inequality).
For any $t_0\in(0,1)$ such that $t_0/\alpha_m-G_m(t_0)<0$, we have
$\P(\thr^{\mathrm{BH}} < t_0) \leq\P( \G(t_0) < t_0/\alpha_m)
\leq\P( \G(t_0)-G_m(t_0) < t_0/\alpha_m-G_m(t_0)).
$
Next, by using Bennett's inequality (see, e.g., Proposition 2.8 in
\cite{Mass2007}) and by letting $h(u)=(1+u)\log(1+u)-u$, for any $u>
0$, we obtain
\[
\P\bigl(\thr^{\mathrm{BH}} < t_0\bigr) \leq\exp\biggl
\{-mG_m(t_0) h \biggl(\frac
{G_m(t_0)-t_0/\alpha_m}{G_m(t_0)} \biggr) \biggr
\}.
\]
Finally, for $t_0\!=\!t_m^{\eps}$, since we have $ G_m(t_m^{\eps
})\!-\!t_m^{\eps}/\alpha_m\! =\! (1\!-\!\pi_{0,m}(1\!-\!\eps))G_m(t_m^{\eps})
\!\geq\!\eps G_m(t_m^{\eps})$, we obtain
(\ref{conc-borneinf}) by using that $h(u)\geq u^2/4$ for any $u>0$.

\subsection{Proofs for the risk $R^T_m$}\label{secotherrisk}

Let us recall that ${R}^T_m$ and ${R}^I_m$ are equal for a
deterministic threshold and thus also for the BFDR threshold. Hence,
Theorem~\ref{main-th} also holds for the risk ${R}^T_m$, and we only
have to prove Theorem~\ref{main-th2}.\looseness=-1

First note that since ${R}^T_m(\thr^{\mathrm{FDR}})={R}^T_m(\thr^{\mathrm{BH}})$, we
can work directly with $\hat{t}^{\mathrm{BH}}_m$. Proving the type I error
bound (\ref{erreurtype1}) can be done similarly: with the same
notation, the type I error can be written conditionally on $H$ as
\begin{eqnarray*}
\E\Biggl( m^{-1} \sum_{i=1}^{m_0(H)}
{\mathbf{1}\bigl\{p_i\leq\thr^{\mathrm{BH}}\bigr\}} \Big| H \Biggr) &
\leq& \E\Biggl( m^{-1} \sum_{i=1}^{m_0(H)}
{\mathbf{1}\{p_i\leq\hat{t}_{m,0}\}} \Big| H \Biggr)
\\
&=& m^{-1} \E(\hat{k}_{m,0} \cond H )
\\
&\leq& \pi_{1,m} \frac{\alpha_m}{1-\alpha_m} + m^{-1} \frac{\alpha
_m}{(1-\alpha_m)^2}
\end{eqnarray*}
by using (\ref{equsansnom}). Hence, (\ref{erreurtype1}) is proved for
the risk ${R}^T_m$.

Next, the proof for bounding the type II error derives essentially from
the following argument, which is quite standard in the multiple testing
methodology; see, for example,~\cite{FZ2006,FDR2009,RV2010,Roq2011}.
Let us denote
\[
\tt= \max\bigl\{t\in[0,1]\dvtx\alpha_m \tG(t)\geq t\bigr\},
\]
where $\tG(t)=m^{-1}(1+\sum_{i=2}^m{\mathbf{1}\{p_i\leq t\}})$
denotes the
empirical c.d.f. of the $p$-values where $p_1$ has been replaced by $0$.
Then, for any realization of the $p$-value family, $p_1\leq\thr^{\mathrm{BH}}$
is equivalent to $p_1\leq\tt$; see, for example, proof of Theorem~2.1
in~\cite{FZ2006} and Section 3.2 of~\cite{Roq2011}.
This entails that the type II error is equal to $\pi_{1,m} (1-\E(F_m(
\tt)))$ [by using\vspace*{1pt} the exchangeability of $(H_i,p_i)_{1\leq i\leq m}$].
Finally, since $\tt\geq\thr^{\mathrm{BH}}$ and $\tt\geq\alpha_m/m$, we
have\vspace*{1pt} $\tt\geq\thr^{\mathrm{FDR}}$. Hence $\pi_{1,m} (1-\E(F_m( \tt)))\leq
\pi_{1,m} (1-\E(F_m( \thr^{\mathrm{FDR}})))$ and bounds (\ref
{erreurtype2-bonf}) and (\ref{erreurtype2}) also hold for the risk ${R}^T_m$.

\section*{Acknowledgments}

We would like to thank Guillaume Lecu\'e and Nicolas Verzelen for
interesting discussions. We are also grateful to anonymous referees, an
Associated Editor, and an Editor for their very helpful comments and
suggestions.

\begin{supplement}
\stitle{Supplement to: On false discovery rate thresholding for
classification under sparsity}
\slink[doi]{10.1214/12-AOS1042SUPP} 
\sdatatype{.pdf}
\sfilename{aos1042\_supp.pdf}
\sdescription{Proofs, additional experiments and supplementary notes
for the present paper.}
\end{supplement}

%

\printaddresses


\begin{thebibliography}{34}

\bibitem{ABDJ2006}
%
\begin{barticle}[mr]
\bauthor{\bsnm{Abramovich},~\bfnm{Felix}\binits{F.}},
\bauthor{\bsnm{Benjamini},~\bfnm{Yoav}\binits{Y.}},
\bauthor{\bsnm{Donoho},~\bfnm{David~L.}\binits{D.~L.}} \AND
\bauthor{\bsnm{Johnstone},~\bfnm{Iain~M.}\binits{I.~M.}}
(\byear{2006}).
\btitle{Adapting to unknown sparsity by controlling the false discovery rate}.
\bjournal{Ann. Statist.}
\bvolume{34}
\bpages{584--653}.
\bid{doi={10.1214/009053606000000074}, issn={0090-5364}, mr={2281879}}
\bptok{imsref}%
\end{barticle}
%
\endbibitem

\bibitem{BH1995}
%
\begin{barticle}[mr]
\bauthor{\bsnm{Benjamini},~\bfnm{Yoav}\binits{Y.}} \AND
\bauthor{\bsnm{Hochberg},~\bfnm{Yosef}\binits{Y.}}
(\byear{1995}).
\btitle{Controlling the false discovery rate: A~practical and powerful approach
to multiple testing}.
\bjournal{J. R. Stat. Soc. Ser. B Stat. Methodol.}
\bvolume{57}
\bpages{289--300}.
\bid{issn={0035-9246}, mr={1325392}}
\bptok{imsref}%
\end{barticle}
%
\endbibitem

\bibitem{BKY2006}
%
\begin{barticle}[mr]
\bauthor{\bsnm{Benjamini},~\bfnm{Yoav}\binits{Y.}},
\bauthor{\bsnm{Krieger},~\bfnm{Abba~M.}\binits{A.~M.}} \AND
\bauthor{\bsnm{Yekutieli},~\bfnm{Daniel}\binits{D.}}
(\byear{2006}).
\btitle{Adaptive linear step-up procedures that control the false discovery
rate}.
\bjournal{Biometrika}
\bvolume{93}
\bpages{491--507}.
\bid{doi={10.1093/biomet/93.3.491}, issn={0006-3444}, mr={2261438}}
\bptok{imsref}%
\end{barticle}
%
\endbibitem

\bibitem{BLS2010}
%
\begin{barticle}[mr]
\bauthor{\bsnm{Blanchard},~\bfnm{Gilles}\binits{G.}},
\bauthor{\bsnm{Lee},~\bfnm{Gyemin}\binits{G.}} \AND
\bauthor{\bsnm{Scott},~\bfnm{Clayton}\binits{C.}}
(\byear{2010}).
\btitle{Semi-supervised novelty detection}.
\bjournal{J. Mach. Learn. Res.}
\bvolume{11}
\bpages{2973--3009}.
\bid{issn={1532-4435}, mr={2746544}}
\bptok{imsref}%
\end{barticle}
%
\endbibitem

\bibitem{BR2009}
%
\begin{barticle}[mr]
\bauthor{\bsnm{Blanchard},~\bfnm{Gilles}\binits{G.}} \AND
\bauthor{\bsnm{Roquain},~\bfnm{{\'E}tienne}\binits{{\'E}.}}
(\byear{2009}).
\btitle{Adaptive false discovery rate control under independence and
dependence}.
\bjournal{J. Mach. Learn. Res.}
\bvolume{10}
\bpages{2837--2871}.
\bid{issn={1532-4435}, mr={2579914}}
\bptok{imsref}%
\end{barticle}
%
\endbibitem

\bibitem{BCFG2010}
%
\begin{barticle}[mr]
\bauthor{\bsnm{Bogdan},~\bfnm{Ma{\l}gorzata}\binits{M.}},
\bauthor{\bsnm{Chakrabarti},~\bfnm{Arijit}\binits{A.}},
\bauthor{\bsnm{Frommlet},~\bfnm{Florian}\binits{F.}} \AND
\bauthor{\bsnm{Ghosh},~\bfnm{Jayanta~K.}\binits{J.~K.}}
(\byear{2011}).
\btitle{Asymptotic {B}ayes-optimality under sparsity of some multiple testing
procedures}.
\bjournal{Ann. Statist.}
\bvolume{39}
\bpages{1551--1579}.
\bid{doi={10.1214/10-AOS869}, issn={0090-5364}, mr={2850212}}
\bptok{imsref}%
\end{barticle}
%
\endbibitem

\bibitem{BGT2008}
%
\begin{bincollection}[mr]
\bauthor{\bsnm{Bogdan},~\bfnm{Ma{\l}gorzata}\binits{M.}},
\bauthor{\bsnm{Ghosh},~\bfnm{Jayanta~K.}\binits{J.~K.}} \AND
\bauthor{\bsnm{Tokdar},~\bfnm{Surya~T.}\binits{S.~T.}}
(\byear{2008}).
\btitle{A comparison of the {B}enjamini--{H}ochberg procedure with some
{B}ayesian rules for multiple testing}.
In \bbooktitle{Beyond Parametrics in Interdisciplinary Research: {F}estschrift
in Honor of {P}rofessor {P}ranab {K}. {S}en}.
\bseries{Inst. Math. Stat. Collect.}
\bvolume{1}
\bpages{211--230}.
\bpublisher{IMS}, \blocation{Beachwood, OH}.
\bid{doi={10.1214/193940307000000158}, mr={2462208}}
\bptok{imsref}%
\end{bincollection}
%
\endbibitem

\bibitem{Chi2007}
%
\begin{barticle}[mr]
\bauthor{\bsnm{Chi},~\bfnm{Zhiyi}\binits{Z.}}
(\byear{2007}).
\btitle{On the performance of {FDR} control: Constraints and a partial
solution}.
\bjournal{Ann. Statist.}
\bvolume{35}
\bpages{1409--1431}.
\bid{doi={10.1214/009053607000000037}, issn={0090-5364}, mr={2351091}}
\bptok{imsref}%
\end{barticle}
%
\endbibitem

\bibitem{DJ2004}
%
\begin{barticle}[mr]
\bauthor{\bsnm{Donoho},~\bfnm{David}\binits{D.}} \AND
\bauthor{\bsnm{Jin},~\bfnm{Jiashun}\binits{J.}}
(\byear{2004}).
\btitle{Higher criticism for detecting sparse heterogeneous mixtures}.
\bjournal{Ann. Statist.}
\bvolume{32}
\bpages{962--994}.
\bid{doi={10.1214/009053604000000265}, issn={0090-5364}, mr={2065195}}
\bptok{imsref}%
\end{barticle}
%
\endbibitem

\bibitem{DJ2006}
%
\begin{barticle}[mr]
\bauthor{\bsnm{Donoho},~\bfnm{David}\binits{D.}} \AND
\bauthor{\bsnm{Jin},~\bfnm{Jiashun}\binits{J.}}
(\byear{2006}).
\btitle{Asymptotic minimaxity of false discovery rate thresholding for sparse
exponential data}.
\bjournal{Ann. Statist.}
\bvolume{34}
\bpages{2980--3018}.
\bid{doi={10.1214/009053606000000920}, issn={0090-5364}, mr={2329475}}
\bptok{imsref}%
\end{barticle}
%
\endbibitem

\bibitem{Efron2008}
%
\begin{barticle}[mr]
\bauthor{\bsnm{Efron},~\bfnm{Bradley}\binits{B.}}
(\byear{2008}).
\btitle{Microarrays, empirical {B}ayes and the two-groups model}.
\bjournal{Statist. Sci.}
\bvolume{23}
\bpages{1--22}.
\bid{doi={10.1214/07-STS236}, issn={0883-4237}, mr={2431866}}
\bptok{imsref}%
\end{barticle}
%
\endbibitem

\bibitem{ET2002}
%
\begin{barticle}[pbm]
\bauthor{\bsnm{Efron},~\bfnm{Bradley}\binits{B.}} \AND
\bauthor{\bsnm{Tibshirani},~\bfnm{Robert}\binits{R.}}
(\byear{2002}).
\btitle{Empirical Bayes methods and false discovery rates for microarrays}.
\bjournal{Genet. Epidemiol.}
\bvolume{23}
\bpages{70--86}.
\bid{doi={10.1002/gepi.1124}, issn={0741-0395}, pmid={12112249}}
\bptok{imsref}%
\end{barticle}
%
\endbibitem

\bibitem{ETST2001}
%
\begin{barticle}[mr]
\bauthor{\bsnm{Efron},~\bfnm{Bradley}\binits{B.}},
\bauthor{\bsnm{Tibshirani},~\bfnm{Robert}\binits{R.}},
\bauthor{\bsnm{Storey},~\bfnm{John~D.}\binits{J.~D.}} \AND
\bauthor{\bsnm{Tusher},~\bfnm{Virginia}\binits{V.}}
(\byear{2001}).
\btitle{Empirical {B}ayes analysis of a microarray experiment}.
\bjournal{J. Amer. Statist. Assoc.}
\bvolume{96}
\bpages{1151--1160}.
\bid{doi={10.1198/016214501753382129}, issn={0162-1459}, mr={1946571}}
\bptok{imsref}%
\end{barticle}
%
\endbibitem

\bibitem{FZ2006}
%
\begin{barticle}[mr]
\bauthor{\bsnm{Ferreira},~\bfnm{J.~A.}\binits{J.~A.}} \AND
\bauthor{\bsnm{Zwinderman},~\bfnm{A.~H.}\binits{A.~H.}}
(\byear{2006}).
\btitle{On the {B}enjamini--{H}ochberg method}.
\bjournal{Ann. Statist.}
\bvolume{34}
\bpages{1827--1849}.
\bid{doi={10.1214/009053606000000425}, issn={0090-5364}, mr={2283719}}
\bptok{imsref}%
\end{barticle}
%
\endbibitem

\bibitem{FDR2009}
%
\begin{barticle}[mr]
\bauthor{\bsnm{Finner},~\bfnm{Helmut}\binits{H.}},
\bauthor{\bsnm{Dickhaus},~\bfnm{Thorsten}\binits{T.}} \AND
\bauthor{\bsnm{Roters},~\bfnm{Markus}\binits{M.}}
(\byear{2009}).
\btitle{On the false discovery rate and an asymptotically optimal rejection
curve}.
\bjournal{Ann. Statist.}
\bvolume{37}
\bpages{596--618}.
\bid{doi={10.1214/07-AOS569}, issn={0090-5364}, mr={2502644}}
\bptok{imsref}%
\end{barticle}
%
\endbibitem

\bibitem{FR2002}
%
\begin{barticle}[mr]
\bauthor{\bsnm{Finner},~\bfnm{H.}\binits{H.}} \AND
\bauthor{\bsnm{Roters},~\bfnm{M.}\binits{M.}}
(\byear{2002}).
\btitle{Multiple hypotheses testing and expected number of type {I} errors}.
\bjournal{Ann. Statist.}
\bvolume{30}
\bpages{220--238}.
\bid{doi={10.1214/aos/1015362191}, issn={0090-5364}, mr={1892662}}
\bptok{imsref}%
\end{barticle}
%
\endbibitem

\bibitem{GBS2009}
%
\begin{barticle}[mr]
\bauthor{\bsnm{Gavrilov},~\bfnm{Yulia}\binits{Y.}},
\bauthor{\bsnm{Benjamini},~\bfnm{Yoav}\binits{Y.}} \AND
\bauthor{\bsnm{Sarkar},~\bfnm{Sanat~K.}\binits{S.~K.}}
(\byear{2009}).
\btitle{An adaptive step-down procedure with proven {FDR} control under
independence}.
\bjournal{Ann. Statist.}
\bvolume{37}
\bpages{619--629}.
\bid{doi={10.1214/07-AOS586}, issn={0090-5364}, mr={2502645}}
\bptok{imsref}%
\end{barticle}
%
\endbibitem

\bibitem{GW2002}
%
\begin{barticle}[mr]
\bauthor{\bsnm{Genovese},~\bfnm{Christopher}\binits{C.}} \AND
\bauthor{\bsnm{Wasserman},~\bfnm{Larry}\binits{L.}}
(\byear{2002}).
\btitle{Operating characteristics and extensions of the false discovery rate
procedure}.
\bjournal{J. R. Stat. Soc. Ser. B Stat. Methodol.}
\bvolume{64}
\bpages{499--517}.
\bid{doi={10.1111/1467-9868.00347}, issn={1369-7412}, mr={1924303}}
\bptok{imsref}%
\end{barticle}
%
\endbibitem

\bibitem{GW2004}
%
\begin{barticle}[mr]
\bauthor{\bsnm{Genovese},~\bfnm{Christopher}\binits{C.}} \AND
\bauthor{\bsnm{Wasserman},~\bfnm{Larry}\binits{L.}}
(\byear{2004}).
\btitle{A stochastic process approach to false discovery control}.
\bjournal{Ann. Statist.}
\bvolume{32}
\bpages{1035--1061}.
\bid{doi={10.1214/009053604000000283}, issn={0090-5364}, mr={2065197}}
\bptok{imsref}%
\end{barticle}
%
\endbibitem

\bibitem{Mass2007}
%
\begin{bbook}[mr]
\bauthor{\bsnm{Massart},~\bfnm{Pascal}\binits{P.}}
(\byear{2007}).
\btitle{Concentration Inequalities and Model Selection}.
\bseries{Lecture Notes in Math.}
\bvolume{1896}.
\bpublisher{Springer}, \blocation{Berlin}.
\bnote{Lectures from the 33rd Summer School on Probability Theory held in
Saint-Flour, July 6--23, 2003, with a foreword by Jean Picard}.
\bid{mr={2319879}}
\bptok{imsref}%
\end{bbook}
%
\endbibitem

\bibitem{NR2011supp}
%
\begin{bmisc}[author]
\bauthor{\bsnm{{N}euvial},~\bfnm{{P}ierre}\binits{P.}} \AND
\bauthor{\bsnm{Roquain},~\bfnm{Etienne}\binits{E.}}
(\byear{2012}).
\bhowpublished{Supplement to ``On false discovery rate thresholding for
classification under sparsity.'' DOI:\doiurl{10.1214/12-AOS1042SUPP}.}
\bptok{imsref}%
\end{bmisc}
%
\endbibitem

\bibitem{Roq2011}
%
\begin{barticle}[mr]
\bauthor{\bsnm{Roquain},~\bfnm{Etienne}\binits{E.}}
(\byear{2011}).
\btitle{Type {I} error rate control for testing many hypotheses: A
survey with
proofs}.
\bjournal{J. SFdS}
\bvolume{152}
\bpages{3--38}.
\bid{issn={2102-6238}, mr={2821220}}
\bptok{imsref}%
\end{barticle}
%
\endbibitem

\bibitem{RV2010}
%
\begin{barticle}[mr]
\bauthor{\bsnm{Roquain},~\bfnm{Etienne}\binits{E.}} \AND
\bauthor{\bsnm{Villers},~\bfnm{Fanny}\binits{F.}}
(\byear{2011}).
\btitle{Exact calculations for false discovery proportion with
application to
least favorable configurations}.
\bjournal{Ann. Statist.}
\bvolume{39}
\bpages{584--612}.
\bid{doi={10.1214/10-AOS847}, issn={0090-5364}, mr={2797857}}
\bptok{imsref}%
\end{barticle}
%
\endbibitem

\bibitem{Sar2002}
%
\begin{barticle}[mr]
\bauthor{\bsnm{Sarkar},~\bfnm{Sanat~K.}\binits{S.~K.}}
(\byear{2002}).
\btitle{Some results on false discovery rate in stepwise multiple testing
procedures}.
\bjournal{Ann. Statist.}
\bvolume{30}
\bpages{239--257}.
\bid{doi={10.1214/aos/1015362192}, issn={0090-5364}, mr={1892663}}
\bptok{imsref}%
\end{barticle}
%
\endbibitem

\bibitem{Sar2008}
%
\begin{barticle}[mr]
\bauthor{\bsnm{Sarkar},~\bfnm{Sanat~K.}\binits{S.~K.}}
(\byear{2008}).
\btitle{On methods controlling the false discovery rate}.
\bjournal{Sankhy\=a}
\bvolume{70}
\bpages{135--168}.
\bid{issn={0972-7671}, mr={2551809}}
\bptok{imsref}%
\end{barticle}
%
\endbibitem

\bibitem{SZG2008}
%
\begin{barticle}[mr]
\bauthor{\bsnm{Sarkar},~\bfnm{Sanat~K.}\binits{S.~K.}},
\bauthor{\bsnm{Zhou},~\bfnm{Tianhui}\binits{T.}} \AND
\bauthor{\bsnm{Ghosh},~\bfnm{Debashis}\binits{D.}}
(\byear{2008}).
\btitle{A general decision theoretic formulation of procedures controlling
{FDR} and {FNR} from a {B}ayesian perspective}.
\bjournal{Statist. Sinica}
\bvolume{18}
\bpages{925--945}.
\bid{issn={1017-0405}, mr={2440399}}
\bptok{imsref}%
\end{barticle}
%
\endbibitem

\bibitem{sawyers08the-cancer}
%
\begin{barticle}[pbm]
\bauthor{\bsnm{Sawyers},~\bfnm{Charles~L.}\binits{C.~L.}}
(\byear{2008}).
\btitle{The cancer biomarker problem}.
\bjournal{Nature}
\bvolume{452}
\bpages{548--552}.
\bid{doi={10.1038/nature06913}, issn={1476-4687}, pii={nature06913},
pmid={18385728}}
\bptok{imsref}%
\end{barticle}
%
\endbibitem

\bibitem{See1968}
%
\begin{barticle}[author]
\bauthor{\bsnm{Seeger},~\bfnm{Paul}\binits{P.}}
(\byear{1968}).
\btitle{A note on a method for the analysis of significances en masse}.
\bjournal{Technometrics}
\bvolume{10}
\bpages{586--593}.
\bptok{imsref}%
\end{barticle}
%
\endbibitem

\bibitem{Sen1999}
%
\begin{barticle}[mr]
\bauthor{\bsnm{Sen},~\bfnm{Pranab~K.}\binits{P.~K.}}
(\byear{1999}).
\btitle{Some remarks on {S}imes-type multiple tests of significance}.
\bjournal{J.~Statist. Plann. Inference}
\bvolume{82}
\bpages{139--145}.
\bnote{Multiple comparisons (Tel Aviv, 1996)}.
\bid{doi={10.1016/S0378-3758(99)00037-3}, issn={0378-3758}, mr={1736438}}
\bptok{imsref}%
\end{barticle}
%
\endbibitem

\bibitem{SW1986}
%
\begin{bbook}[mr]
\bauthor{\bsnm{Shorack},~\bfnm{Galen~R.}\binits{G.~R.}} \AND
\bauthor{\bsnm{Wellner},~\bfnm{Jon~A.}\binits{J.~A.}}
(\byear{1986}).
\btitle{Empirical Processes with Applications to Statistics}.
\bpublisher{Wiley}, \blocation{New York}.
\bid{mr={0838963}}
\bptok{imsref}%
\end{bbook}
%
\endbibitem

\bibitem{Storey2002}
%
\begin{barticle}[mr]
\bauthor{\bsnm{Storey},~\bfnm{John~D.}\binits{J.~D.}}
(\byear{2002}).
\btitle{A direct approach to false discovery rates}.
\bjournal{J. R. Stat. Soc. Ser. B Stat. Methodol.}
\bvolume{64}
\bpages{479--498}.
\bid{doi={10.1111/1467-9868.00346}, issn={1369-7412}, mr={1924302}}
\bptok{imsref}%
\end{barticle}
%
\endbibitem

\bibitem{Storey2003}
%
\begin{barticle}[mr]
\bauthor{\bsnm{Storey},~\bfnm{John~D.}\binits{J.~D.}}
(\byear{2003}).
\btitle{The positive false discovery rate: A {B}ayesian interpretation
and the
{$q$}-value}.
\bjournal{Ann. Statist.}
\bvolume{31}
\bpages{2013--2035}.
\bid{doi={10.1214/aos/1074290335}, issn={0090-5364}, mr={2036398}}
\bptok{imsref}%
\end{barticle}
%
\endbibitem

\bibitem{TLD1998}
%
\begin{barticle}[mr]
\bauthor{\bsnm{Tamhane},~\bfnm{Ajit~C.}\binits{A.~C.}},
\bauthor{\bsnm{Liu},~\bfnm{Wei}\binits{W.}} \AND
\bauthor{\bsnm{Dunnett},~\bfnm{Charles~W.}\binits{C.~W.}}
(\byear{1998}).
\btitle{A generalized step-up-down multiple test procedure}.
\bjournal{Canad. J. Statist.}
\bvolume{26}
\bpages{353--363}.
\bid{doi={10.2307/3315516}, issn={0319-5724}, mr={1648451}}
\bptok{imsref}%
\end{barticle}
%
\endbibitem

\end{thebibliography}
\end{document}